\documentclass[dvipdfmx]{emulateapj}

\usepackage{url}

\shortauthors{Yasui et al.}

\begin{document}

\title{Low-metallicity Young Clusters in the Outer
Galaxy. III. S\lowercase{h} 2-127}

\author{Chikako Yasui\altaffilmark{1}, Naoto 
Kobayashi\altaffilmark{2, 3, 4}, 
Masao Saito\altaffilmark{5, 6}, Natsuko Izumi\altaffilmark{7, 8},
and Warren Skidmore\altaffilmark{9}}

\affiliation{$^1$National Astronomical Observatory of Japan,
California Office, 
100 W. Walnut St., Suite 300, Pasadena, CA 91124, USA \\
{ck.yasui@gmail.com} \\
$^2$Institute of Astronomy, School of Science, University of Tokyo,
2-21-1 Osawa, Mitaka, Tokyo 181-0015, Japan \\
$^3$Kiso Observatory, Institute of Astronomy, School of Science,
University of Tokyo, 10762-30 Mitake, Kiso-machi, Kiso-gun, Nagano
397-0101, Japan \\
$^4$Laboratory of Infrared High-resolution spectroscopy (LIH), Koyama
Astronomical Observatory, Kyoto Sangyo University, Motoyama, Kamigamo,
Kita-ku, Kyoto 603-8555, Japan \\
$^5$ National Astronomical Observatory of Japan,
2-21-1 Osawa, Mitaka, Tokyo 181-8588, Japan \\
$^6$The Graduate University of Advanced Studies 
(SOKENDAI), 2-21-1 Osawa, Mitaka, Tokyo 181-8588, Japan \\
$^7$College of Science, Ibaraki University, 2-1-1 Bunkyo, Mito, Ibaraki
310-8512, Japan \\
$^8$Institute of Astronomy and Astrophysics, Academia Sinica, No. 1,
Section 4, Roosevelt Road, Taipei 10617, Taiwan \\
$^9$Thirty Meter Telescope International Observatory, 137 W Walnut Ave,
Monrovia, CA 91016, USA}


%
\begin{abstract}
In deep near-infrared imaging of the low-metallicity (${\rm [O/H]}=-0.7$
dex) \ion{H}{2} region Sh 2-127 (S127) with Subaru/MOIRCS, we detected
two young clusters with 413 members (S127A) in a slightly extended
\ion{H}{2} region and another with 338 members (S127B) in a compact 
\ion{H}{2} region.
The limiting magnitude was $K=21.3$ mag (10$\sigma$), corresponding to a
mass detection limit of $\sim$0.2 $M_\odot$.
These clusters are an order of magnitude larger than previously studied
young low-metallicity clusters and larger than the majority of solar
neighborhood young clusters.
Fits to the {\it K}-band luminosity functions indicate very young
cluster ages of 0.5 Myr for S127A and 0.1--0.5 Myr for S127B, consistent
with the large extinction (up to $A_V\simeq20$ mag) from thick molecular
clouds and the presence of a compact \ion{H}{2} region and class I
source candidates, and suggest that the initial mass function (IMF) of
the low-metallicity clusters is indistinguishable from typical solar
neighborhood IMFs.
Disk fractions of $28\% \pm 3\%$ for S127A and $40\% \pm 4\%$ for S127B
are significantly lower than those of similarly aged solar neighborhood
clusters ($\sim$50\%--60\%).
The disk fraction for S127B is higher than those of previously studied
low-metallicity clusters ($<$30 \%), probably due to S127B's age.
This suggests that a large fraction of very young stars in
low-metallicity environments have disks, but the disks are lost on a
very short timescale.
These results are consistent with our previous studies of
low-metallicity star-forming regions, suggesting that a solar
neighborhood IMF and low disk fraction are typical characteristics for
low-metallicity regions, regardless of cluster scales.

\end{abstract}




\section{INTRODUCTION} \label{sec:intro}
\setcounter{footnote}{9}

Considering that the only elements in existence at the beginning of our
universe were H and He (and an insignificant amount of Li), and that the
present chemical composition of the universe is the result of metal
pollution due to elemental synthesis inside stars and the influence of
supernovae explosions, exploring the physical processes involved with star
formation under different metallicity environments is of great interest.
Although even at present, metals only account for 2 \% of the mass in
our solar system and the local universe, they can be critical factors
for star and planet formation;
metals strongly affect heating and cooling related to radiative transfer
in star formation processes.  Also in planet formation processes, dust
forms planet cores despite being only a very small mass fraction in
disks (only $\sim$1 \%).
Therefore, observations of the metallicity dependence of these processes
can put strong constraints on theories of star and planet formation
\citep{{Chruslinska2020},{Ercolano2010}}.

If we observe the universe at redshifts $\ge 1$, we see the early
universe, where metallicity is lower than at present.
Even in our Galaxy, there is a wide range of
metallicity, from $-$1 to $+$0.5 dex \citep[e.g.,][]{Rudolph2006}. 
However, to see the full details of the star formation processes, it is
necessary to make spatially resolved observations down to the substellar
mass regime ($\sim$0.1 $M_\odot$).
This enables us to compare results obtained for low-metallicity
environments with those in the solar neighborhood on the same basis.
Therefore, we are focusing on our Galaxy, the only region where
observations down to the substellar mass regime can be conducted with
existing telescopes and instruments.
Although the Large Magellanic Cloud (LMC; $D\sim 50$ kpc) and Small
Magellanic Cloud (SMC; $D\sim 620$ kpc) are well known as extragalactic
objects that are relatively close to the Sun, the mass detection limit
can only reach about 1 $M_\odot$ with current facilities due to the
larger distances compared to within our Galaxy ($D\sim 10$--20 kpc).
Not until the availability of the extremely large telescopes that will a
detection limit for extragalactic objects reaching the substellar limit
finally be achieved \citep{{Yasui2020},{Leschinski2020}}. 
The solar neighborhood is an environment where the metallicity is
relatively high within our Galaxy, as well as relative to the local
region ($\sim$100--200 pc) \citep{Santos2008}.  To examine the possible
dependence on metallicity of the star formation process, we selected
targets with low-metallicity environments ($\sim$$-$1 dex) as a first
step.  A metallicity of $\sim$$-$1 dex corresponds to a redshift of
$z\sim2$ \citep{Pettini1997}, when the age of the universe was about 3.3
Gyr, and to the thick-disk formation phase in our Galaxy
\citep{Brewer2006}.  The elucidation of star and planet formation
processes under these circumstances is the purpose of this paper.

Prior to our campaign, there have only been a few published studies of
star-forming regions in low-metallicity environments:
\citet{Kobayashi2000}, \citet{Santos2000}, \citet{Snell2002},
\citet{Brand2007}, and \citet{Kobayashi2008}.
The mass detection limits of previous studies have been surpassed in
the studies presented here and other papers in our series, reaching
down to the substellar mass regime ($\sim$0.1 $M_\odot$): Digel Cloud 2
\citep{{Yasui2006}, {Yasui2008}}, S207 \citep{Yasui2016_S207}, and S208
\citep{Yasui2016_S208}.
The first example is the Cloud 2 clusters, which are associated with a
molecular cloud \citep{Digel1994} located in the extreme outer Galaxy
with a Galactrocentric distance ($R_G$) of $\simeq$19 kpc in the
direction of $(l, b) = (137.75, -1.0)$.  The Cloud 2 clusters are beyond
the Norma-Cygnus (outer) arm (previously considered the outermost arm)
but may be located in the recently discovered extension of the
Scutum-Centaurus arm \citep{Sun2015}.  The star-forming activities in
this region were identified by \citet{Kobayashi2000}, and
\citet{Kobayashi2008} identified two young clusters, the Cloud 2-N and
-S clusters.  The metallicity of this region is estimated to be
$\sim$$-$0.7 dex from an associated B-type star \citep{{Smartt1996},
{Rolleston2000}} and the Digel Cloud 2 molecular cloud itself
\citep{Lubowich2004}.  From near-infrared (NIR) imaging,
\citet{Yasui2008} identified 72 and 66 cluster members in the Cloud 2-N
and -S clusters, respectively, with a mass detection limit of $\sim$0.1
$M_\odot$ and estimated the ages as $\sim$0.5 Myr for both clusters.

This paper is one in a series of papers in which we present deep NIR
imaging observations of low-metallicity young clusters in the Galaxy and
analysis of those images. These clusters were selected using the
following:
i) the Sharpless catalog (a list of H$\alpha$-selected bright \ion{H}{2}
regions; \citealt{Sharpless1959}), 
ii) the region associated with clusters having a significant number of
cluster members, and 
iii) an oxygen metallicity ${\rm [O/H]} \le -0.5$ dex, assuming a
solar abundance of $12 + \log {\rm (O/H)} = 8.73$ \citep{Asplund2009}.
Among $\sim$10 selected regions, we first presented the results of S207
\citep[][hereafter Paper I]{Yasui2016_S207} and S208 \citep[hereafter
Paper II]{Yasui2016_S208}, which are two of the lowest-metallicity
\ion{H}{2} regions (each with ${\rm [O/H]} \simeq$ $-$0.8 dex) in the
direction of $(l, b) = (151. 2, 2.13)$ in Galactic coordinates.
With NIR deep imaging, we identified one cluster in each region, having
73 and 89 cluster members in S207 and S208, respectively, with a
mass detection limit of $\lesssim$0.1 $M_\odot$ for S207 and $\sim$0.2
$M_\odot$ for S208.
From the {\it K}-band luminosity function (KLF) fitting of the clusters,
S207 and S208 are likely to be located at $D=4$ kpc from the Sun, which
suggests that two regions are located in the interarm region between
the Cygnus and Perseus arms \citep{Vallee2020}.
The fitting also suggested that S207 is at the end of the embedded
cluster phase ($\sim$2--3 Myr), while the S208 cluster is very young
($\sim$0.5 Myr).

From the results of four low-metallicity star-forming clusters 
(two Cloud 2 clusters, the S207 cluster, and the S208 cluster), the
initial mass function (IMF) 
in low-metallicity clusters is suggested to be consistent with the
typical IMF obtained in clusters with solar metallicity.
In contrast, the estimated disk fractions in low-metallicity clusters
are lower than those in the solar neighborhood, suggesting that the
lifetime of protoplanetary disks in such environments is shorter than in
clusters with solar metallicity \citep{{Yasui2009}, {Yasui2010}}.
However, the previous low-metallicity clusters studied are relatively
small, with the number of cluster members ($N_{\rm stars}$) $<$100,
and it is important to see if these results are also valid in larger
clusters, e.g. $N_{\rm stars}$ $>$100.
For example, in a solar metallicity environment, it has been found that
the IMF is different between starburst clusters with $N_{\rm stars}$ 
$>$$10^4$ and more common clusters with $N_{\rm stars}$ 
$\sim$10--$10^3$ (see Section~\ref{sec:IMF_age}), and the disk
fraction in more massive or denser clusters is suggested to be low
(see Section~\ref{sec:DF}).

In this paper, we present the results for our third target, Sh 2-127
(S127), which is a low-matallicity star-forming region in the Galaxy
with ${\rm [O/H]} \simeq -0.7$ dex, as for previous targets, located
between the Cygnus and Scutum-Crux arms.
However, within S127, we found two star-forming clusters with
$\sim$300--400 cluster members ($N_{\rm stars}$), more than an order of magnitude
higher than in previous targets.
This paper is organized as follows. Section~2 describes previous studies
of S127, focusing on studies of star-forming activities in S127 using
multiwavelength data, e.g., mid-infrared (MIR) data from the 
Wide-field Infrared Survey Explorer ({\it WISE}),
NIR data from the Two Micron All Sky Survey (2MASS),
H$\alpha$ data from 
the Isaac Newton Telescope Photometric H-Alpha Survey (IPHAS),
and radio continuum emission from the NRAO VLA Sky Survey (NVSS).
Section~3 describes our Subaru
Multi-Object InfraRed Camera and Spectrograph (MOIRCS)
deep $JHK_S$ images and data reduction. 
Section~4 describes the results for the star-forming clusters in S127. 
In Section~5, we discuss the basic cluster parameters, such as
cluster scale, age, IMF, and disk fraction, of the S127 clusters.
Finally, in Section~6, we discuss the IMF 
the low-metallicity environment.

\section{S127} \label{sec:S127} 
In this section, the properties of the target star-forming region, S127,
are summarized.  In Table~\ref{tab:targets}, we summarize the properties
from previous works, including coordinates, distance, oxygen abundance,
and metallicity.  We also show the large-scale NIR and MIR pseudocolor
and H$\alpha$ images of S127 in Figure~\ref{fig:3col_2MASS_WISE}.

\subsection{Basic Properties from the Literature} \label{sec:properties}

The region S127 is located at $(l, b) = (96.287^\circ, +2.594^\circ)$ on
the Galactic plane with coordinates of $(\alpha_{\rm 2000.0},
\delta_{\rm 2000.0}) = (21^{\rm h} 28^{\rm m} 41.6^{\rm s}, +54^\circ
37' 00'')$ from the SIMBAD database\footnote{This research has made use
of the SIMBAD database, operated at the Centre de Donn\'ees
Astronomiques de Strasbourg, France.} \citep{Wenger2000}.  It has an
extended \ion{H}{2} region traced by H$\alpha$ \citep{Sharpless1959} and
radio continuum \citep{Fich1993} emission.  Strong MIR emission is
detected with IRAS, IRAS 21270+5423 in the IRAS Point Source Catalog
(\citealt{Beichman1988}; see the large red plus sign in
Figure~\ref{fig:3col_2MASS_WISE}) and IRAS X2127+544 in the IRAS
Small-Scale Structure Catalog \citep{Helou1988}.  CO emission is
reported in e.g., \citet{Blitz1982} and \citet{Wouterloot1989}, and the
results of high-resolution CO observations are shown in
\citet{Brand2001}.  A star-forming cluster is identified by
\citet{Bica2003} as [BDS2003]24 using 2MASS images, with a center of
$(\alpha_{\rm 2000.0}, \delta_{\rm 2000.0}) = (21^{\rm h} 28^{\rm m}
43^{\rm s}, +54^\circ 37' 03'')$ and angular dimensions of $1\farcm9
\times 1\farcm1$.  The photometric distance, which is determined from
spectroscopic and photometric observations, is estimated to be 9.7 kpc
for the O8V-type star \citep{Chini1984}, ALS 18695 (see the small red
plus sign in Figure~\ref{fig:3col_2MASS_WISE}).  The kinematic distance
is also estimated at $\simeq$10 kpc using the radial velocities of
$V_{\rm LSR} \sim -95$ km s$^{-1}$ derived in various observations:
$V_{\rm LSR} = -94.5$ km s$^{-1}$ for CO observations by
\citet{Blitz1982}, $V_{\rm LSR} = -98.9$ km s$^{-1}$ for the H$\alpha$
Fabry--Perot observation by \citet{Fich1990}, $V_{\rm LSR}=-92.09$ km
s$^{-1}$ for CO observations by \citet{Wouterloot1989}, $V_{\rm
LSR}=-94.7$ km s$^{-1}$ for Fabry--Perot observations of the \ion{H}{2}
region by \citet{Caplan2000}, and $V_{\rm LSR}=-94.7$ and $-$92.98 km
s$^{-1}$ for $^{12}$CO line data by FCRAO and \ion{H}{1} data by CGPS,
respectively \citep{Foster2015}.
According to \citet{Foster2015}, the most recent derivation, the
estimated distance is $9.97\pm 1.73$ kpc, which is consistent with the
kinematic distance.  Assuming that the Galactocentric distance of the
Sun is $R_\odot = 8.0$ kpc, the distance corresponds to $R_G \simeq
13.5$ kpc.  Figure~\ref{fig:Gal} shows a top view of the Galaxy with
S127 and spiral arms.  The location of S127 is shown by the filled
circle, while the spiral arms are shown with different colors, e.g., red
and cyan for the Norma-Cygnus (outer) and Scutum-Crux arms,
respectively.  The figure shows that S127 is located between the Cygnus
arm and the extension of the Scutum-Crux arm.

The oxygen abundances (O/H) of S127 have been estimated from optical and
far-IR (FIR) observations.  \citet{Vilchez1996} estimated the abundance
for eight \ion{H}{2} regions by measuring optical emission line fluxes
in spectroscopic observations, while \citet{Caplan2000} measured optical
emission line fluxes for 36 \ion{H}{2} regions based on Fabry--Perot
observations.\footnote{\citet{Deharveng2000} subsequently derived the
abundances using data presented by \citet{Caplan2000}.}
\citet{Rudolph1997} estimated the abundance for five \ion{H}{2} regions
by measuring FIR emission line fluxes with the Kuiper Airborne
Observatory, while \citet{Peeters2002} measured line fluxes between 2.3
and 196 $\mu$m mainly for IRAS point sources in 45 (compact) \ion{H}{2}
regions based on Infrared Space Observatory (ISO)
spectroscopy.\footnote{\citet{Martin-Hernandez2002} subsequently derived
the oxygen abundance using data by \citet{Peeters2002}.}
\citet{Rudolph2006} reanalyzed the elemental abundances of 117
\ion{H}{2} regions with updated physical parameters.  Among them, the
oxygen abundances of S127 are estimated to be $12 + \log ({\rm O/H}) =
7.68$--8.46 ($8.15^{+0.12}_{-0.17}$, $8.20^{+0.15}_{-0.23}$,
$8.20^{+0.17}_{-0.29}$, $7.68^{+0.17}_{-0.29}$, and
$8.46^{+0.30}_{-1.65}$, using the data by \citealt {Vilchez1996},
\citealt{Rudolph1997}, \citealt{Caplan2000}, \citealt{Peeters2002}, and
\citealt{Rudolph2006}, respectively).\footnote{\citet{Rudolph2006}
collected some estimations from several independent studies for various
positions: $8.15^{+0.12}_{-0.17}$ and $8.20^{+0.17}_{-0.29}$ both within
S127, $8.20^{+0.15}_{-0.23}$ for S127A, $8.46^{+0.30}_{-1.65}$ for
S127B, and $7.68^{+0.17}_{-0.29}$ for IRAS 21270+5423.}
This corresponds to a metallicity of ${\rm [O/H]} \simeq -0.7$ dex
assuming the solar abundance of $12 + \log ({\rm O/H}) = 8.73$
\citep{Asplund2009}.  The electron temperatures ($T_{\rm e}$) are also
sensitive indicators of the abundances, with higher temperatures for
lower abundances \citep{Shaver1983}. The estimated temperatures are very
high for S127, $\sim$11000 K, e.g., 10500$\pm$820 K from
\citet{Scaife2008} and 11428$\pm$305 K from \citet{Balser2011}.
They are some of the highest temperatures among \ion{H}{2} regions in
our Galaxy, suggesting that S127 is a very low-metallicity region.
According to the relationship between the electron temperatures and
oxygen abundances by \citet{Shaver1983} of $12 + \log {\rm (O/H)} = 9.82
- 1.49 \ T_{\rm e} / 10^4$, the temperature of S127 ($\sim$11000 K)
suggests an oxygen abundance of 8.2, which is consistent with the
abundance estimation described earlier.

\subsection{Star-forming Activities} \label{sec:SFinS127}

\begin{figure*}[h]
 \begin{center}
  \vspace{10em}
 \includegraphics[width=9cm]{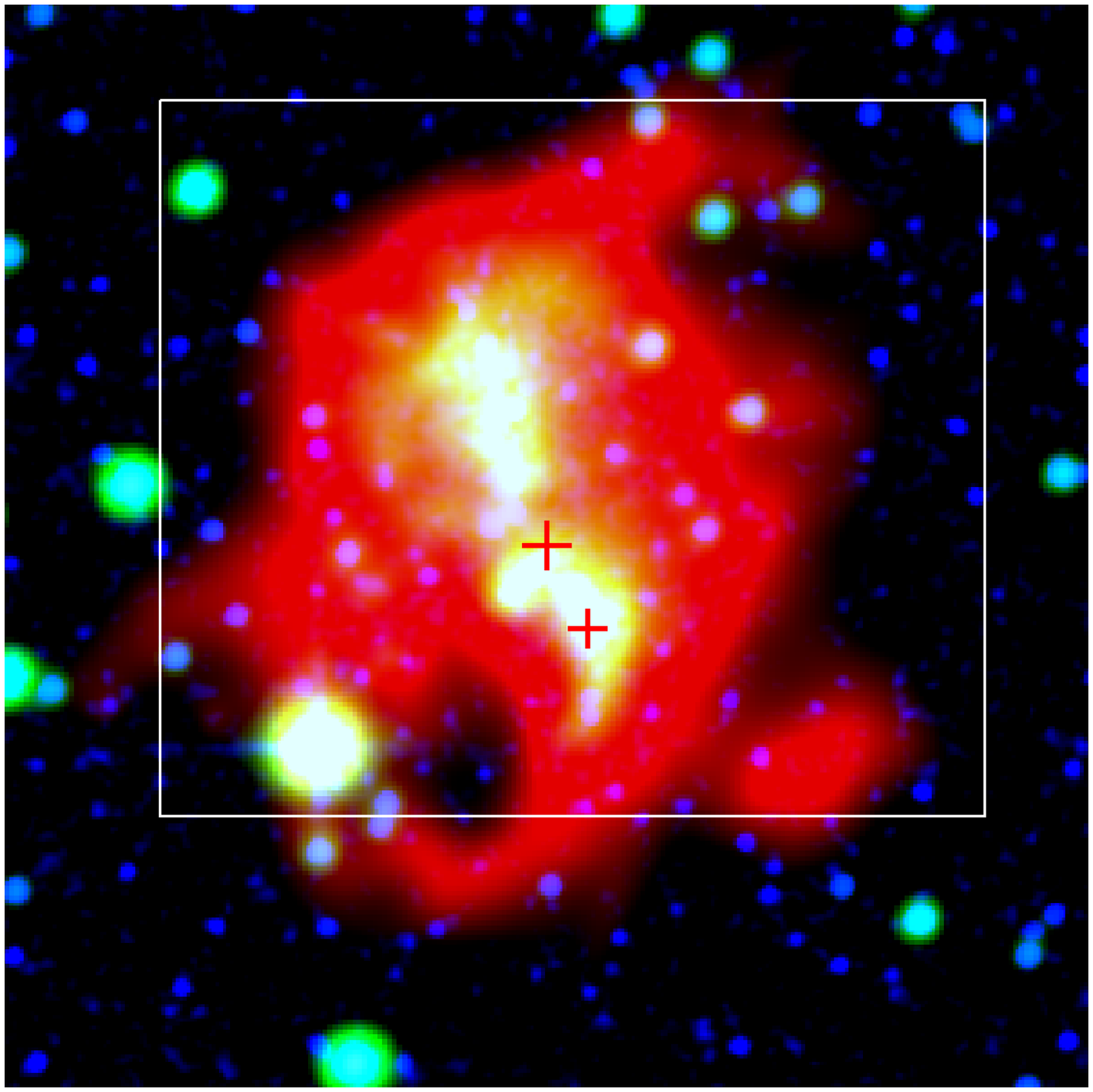}
 \includegraphics[width=9cm]{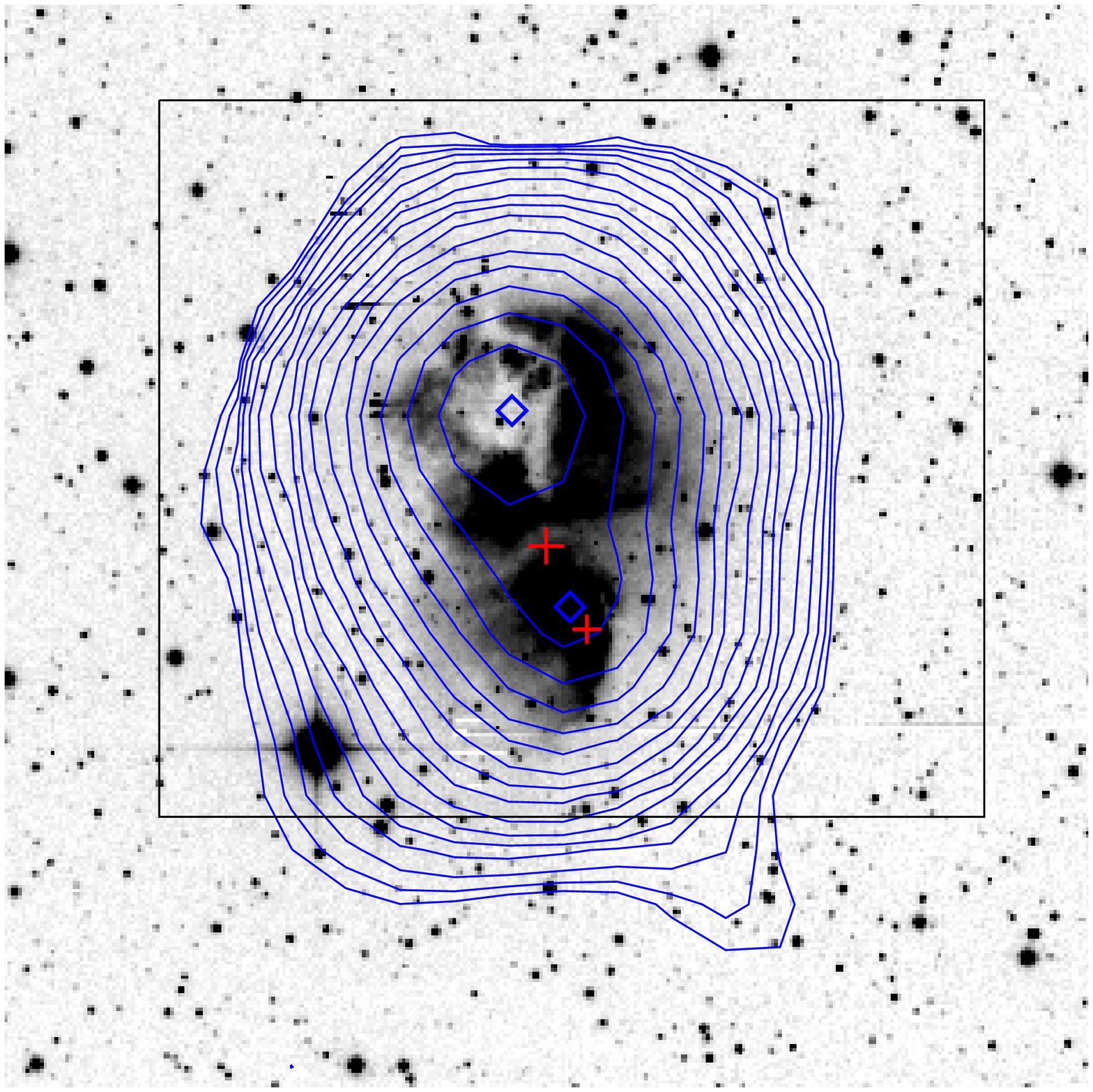}
  \vspace{-5em}
 
\caption{Pseudocolor (top) and H$\alpha$ (bottom) images  of S127
with a wide field of view of $5'\times 5'$ centered at $(\alpha_{\rm
2000.0}, \delta_{\rm 2000.0}) = (21^{\rm h} 28^{\rm m} 42^{\rm s},
+54^\circ 36' 51'')$ in equatorial coordinates and $(l, b) =
(96.286^\circ, +2.592^\circ$) in Galactic coordinates, which is the
 coordinate of IRAS 21270+5423.
North is up, and east is to the left. 
 The $1\arcmin$ corresponds to 2.9 pc for the distances of S127.
Top: the image is produced by combining the 2MASS
 $K_S$-band (2.16 $\mu$m; blue), {\it WISE} band 1 (3.4 $\mu$m; green),
 and {\it WISE} band 3 (12 $\mu$m; red).  The large red plus
 sign shows the IRAS point source, while the
 small red plus sign shows the bright stars in the
 optical bands, ALS 18695.  The white box shows the location and size of
 the MOIRCS field of view. 
 Bottom: IPHAS H$\alpha$ image of S127 shown in gray scale. 
 The 1.4 GHz radio continuum emission by NVSS is also shown with blue
 contours.  The contours are plotted at 1 mJy ${\rm beam}^{-1} \times
 2^0, 2^{-1/2}, 2^1$, ... .  The blue diamonds show the NVSS radio point
 sources, NVSS 212841+543634 and NVSS 212843+543728.  The red plus
 symbols are the same as those in the top panel, while the black box is
 the same as the white box in the top panel.}
 \label{fig:3col_2MASS_WISE}
 \end{center}
\end{figure*}

\begin{figure}[h]
\begin{center}
  \includegraphics[width=10cm]{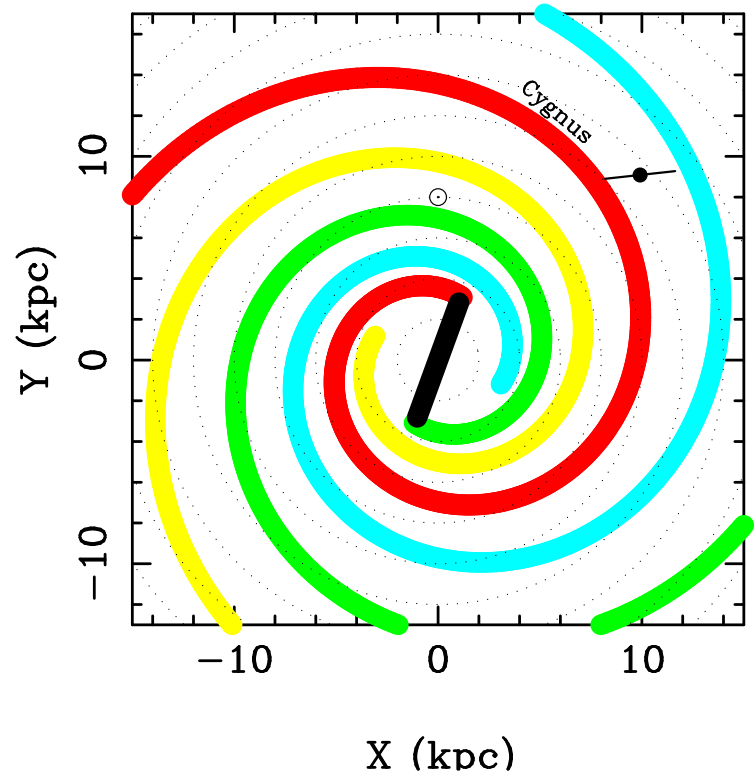}
\caption{Top view of the Milky Way galaxy, showing S127 in relation to
the spiral arms.  The filled circle shows S127 at a distance of $D=9.97
\pm 1.73$ kpc \citep{Foster2015} from the sun. The Sun is shown at a
Galactocentric distance of 8 kpc by a circled dot. Spiral arms from
\citet{Vallee2005} are shown with different colors (red, yellow, green,
and cyan for the Norma-Cygnus, Perseus, Sagittarius-Carina, and
Scutum-Crux arms, respectively).}  \label{fig:Gal}
\end{center}
\end{figure}

The top panel of Figure~\ref{fig:3col_2MASS_WISE} shows an NIR and MIR
pseudocolor image of S127 with a wide field of view ($5\arcmin \times
5\arcmin$) centered at IRAS 21270+5423, shown with the large red plus
sign.  The figure is produced by combining the 2MASS
\citep{Skrutskie2006} {\it K}$_S$-band (2.16\,$\mu$m; blue), {\it WISE}
\citep{Wright2010} band 1 (3.4 $\mu$m; green), and {\it WISE} band 3 (12
$\mu$m; red) images.
Two components can be seen to the north and south of the IRAS source
(large red plus sign) in {\it WISE} band 3, whose emission is mainly
from polycyclic aromatic hydrocarbon (PAH) emission, tracing
photodissociation regions around \ion{H}{2} regions.
The bottom panel of Figure~\ref{fig:3col_2MASS_WISE} shows an H$\alpha$
image from IPHAS \citep{Drew2005} in gray scale and the 1.4 GHz radio
continuum from NVSS \citep{Condon1998} in blue contours.
The overall distribution of H$\alpha$ and radio continuum emissions that
trace the photoionized \ion{H}{2} region is consistent with that of 12
$\mu$m features. There are two NVSS radio sources in this field, NVSS
212841+543634 and NVSS 212843+543728, indicated by blue diamonds, which
are located in the centers of the northern and southern components,
but the distribution of the radio continuum is not divided into two
components at the north and south of the IRAS source due to the
relatively low spatial resolution of NVSS (45$\arcsec$).  The images of
radio continuum emissions with higher spatial resolution (5$\farcs$5) by
\citet{Rudolph1996} show the two components, and there are two NVSS
radio sources in this field, NVSS 212841+543634 and NVSS 212843+543728,
indicated by blue diamonds, which are located in the center of the
northern and southern component.  \citet{Rudolph1996} refer to northern
and southern regions as S127A and S127B, respectively.
They estimated the properties of the \ion{H}{2} regions (S127A and
S127B) separately.
Their results showed that S127A is an extended \ion{H}{2} region (1.7 pc
diameter), while S127B is a compact \ion{H}{2} region (0.5 pc diameter).
\citet{Rudolph1996} also pointed out that the two regions have cometary
shapes, which may be due to pressure confinement of the expanding
ionized gas, the so-called champagne flow.  \citet{Brand2001} presented
high-resolution CO observations.
They showed northern and southern CO complexes around the S127A and
S127B \ion{H}{2} regions.  The distribution of CO emissions correlates
fairly well with that of radio continuum emissions.  The northern CO
complex around S127A obscures most of the optical emission,
suggesting that the CO complex is located in the foreground.
This is also seen as reduced emission (the gray area within S127A) in
the H$\alpha$ image (bottom panel of Figure~\ref{fig:3col_2MASS_WISE})
around the northern NVSS source.

The compact \ion{H}{2} region (S127B) is located at the eastern edge of
a patch of obscuration, which corresponds to a peak of the southern CO
complex.  This suggests that the \ion{H}{2} region lies on the near side
of the molecular complex.  This is also seen in the H$\alpha$ image
(bottom panel of Figure~\ref{fig:3col_2MASS_WISE}) to the west of the
southern NVSS source marked by the abrupt edge.
Because ALS 18695 is located around the center of the southern
\ion{H}{2} region in the vicinity of the southern NVSS source, the
O-type star should be the exciting source of the southern \ion{H}{2}
region.  The spectral type of ALS 18695 (O8V) is consistent with the
results of \citet{Rudolph1996} that the spectral type of a single
zero-age main-sequence (ZAMS) star that could provide a flux of ionizing
photons of the S127B \ion{H}{2} region would be O8.5.

\section{OBSERVATIONS AND DATA REDUCTION}

\subsection{Subaru MOIRCS JHK Imaging} \label{sec:obs_MOIRCS}

Using the same instrumental setup described in Papers I and Paper II,
MOIRCS \citep{{Ichikawa2006},{Suzuki2008}} was used on the 8.2 m Subaru
telescope to obtain deep $JHK_S$ images with the Mauna Kea Observatories
(MKO) NIR filters \citep{{Simons2002},{Tokunaga2002}} over a $3\farcm5
\times 4'$ field at $0\farcs117$ pixel$^{-1}$.

The long-exposure observations described in this paper were performed on
2006 September 2 UT.  Observing conditions were photometric, and the
seeing was excellent ($\sim$$0\farcs35$--$0\farcs45$) throughout the night.
For the long-exposure images, individual exposure times for the $J$,
$H$, and $K_S$ bands were 150, 20, and 30 s, respectively, while the total
integration times are 1350, 1080, and 1080 s for the {\it J}, {\it H},
and $K_S$ bands, respectively.  For counts over 20,000 ADU, the detector
output linearity is not guaranteed.  To ensure accurate flux calibration
for the brightest targets in the cluster, we also obtained
short-exposure images on 2007 November 22 UT.  The exposure time for
individual short-exposure images is 13 s, and the total integration time
is 52 s for all bands.  The whole \ion{H}{2} region described in
Section~\ref{sec:SFinS127} is covered by one chip
($\sim$$3\farcm5\times4\farcm0$, hereafter the ``S127
frame''; see the white and black boxes in
Figure~\ref{fig:3col_2MASS_WISE}), whose center is at $\alpha_{\rm 2000}
= 21^{\rm h} 28^{\rm m} 41.2^{\rm s}$, $\delta_{\rm 2000} = +54^\circ
37' 16\farcs2$.
For the background subtraction, to avoid the nebulosity of S127, the
telescope was nodded by $3\farcm5$ north and south (equal to the field
of view for one detector chip) in addition to 10$''$ dithering, the two
chips being alternately directed to the field, continuously observing
the field.  After the long-exposure images were obtained, one of the
MOIRCS science grade detectors was exchanged for an engineering grade
detector.  Although this engineering grade detector was used while
gathering short-exposure images, we only included observations obtained
with the science grade detector in our analysis.
The $3\farcm5$ sky region north of S127, whose images are obtained with
a science grade detector in both long- and
short-exposure images, is defined as the ``sky frame.'' 
We summarize the details of the observations in Table~\ref{tab:LOG}.

\subsection{Data Reduction and Photometry} \label{sec:Data} 

\begin{figure}[h]
 \begin{center}
  \vspace{20em}
  \hspace{-10em}
  \includegraphics[width=19.5cm]{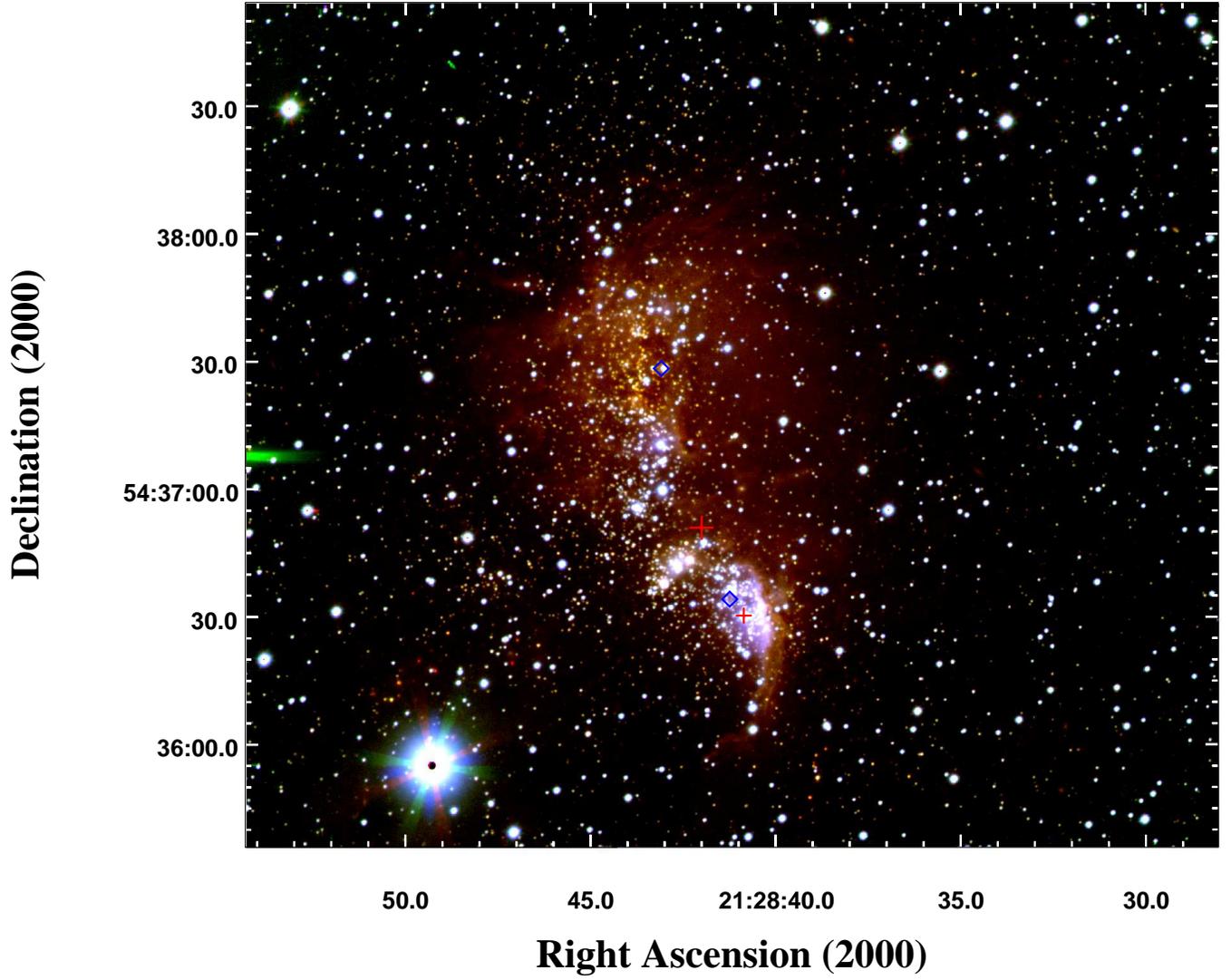}
  \vspace{-20em}
 \caption{Pseudocolor image of S127 produced by
 combining the $J$- (1.26 $\mu$m), $H$- (1.64 $\mu$m), and $K_S$-band
 (2.15 $\mu$m) MOIRCS images from 2006 September. The
 equatorial coordinates of the center of the image are $\alpha_{\rm
 2000} = 21^{\rm h} 28^{\rm m} 41.2^{\rm s}$, $\delta_{\rm 2000} =
 +54^\circ 37' 16.2''$.  The field of view of $\sim$$3\farcm5 \times
 4\arcmin$ is shown with white and black boxes in
 Figure~\ref{fig:3col_2MASS_WISE}, and the symbols are
 the same as in Figure~\ref{fig:3col_2MASS_WISE}. }
 \label{fig:3col_S127}
 \end{center}
\end{figure}

All data were reduced with IRAF\footnote{IRAF is distributed by the
National Optical Astronomy Observatories, which are operated by the
Association of Universities for Research in Astronomy, Inc., under
cooperative agreement with the National Science Foundation.}  using the
procedure described in Papers I and II: flat fielding,
bad-pixel correction, median-sky subtraction, image shifts with
dithering offsets, and image combination.  For the flat fielding, sky
flats were used that were made with MOIRCS data in the
closest run obtained from the SMOKA\footnote{SMOKA is the
Subaru--Mitaka--Okayama--Kiso Archive System operated by the Astronomy
Data Center, National Astronomical Observatory of Japan.} data archive.
Before any image combination, the MOIRCS image reduction package
MCSRED\footnote{\url{http://www.naoj.org/staff/ichi/MCSRED/mcsred\_e.html}}
was used to correct for image distortion using the process described in
Papers I and II.  Figure~\ref{fig:3col_S127} shows a pseudocolor
image of S127 constructed by combining the $J$ (1.26 $\mu$m; blue), $H$
(1.64 $\mu$m; green), and $K_S$ (2.15 $\mu$m; red) long-exposure images.

For the long-exposure images, photometry with point spread function
(PSF) fitting using IRAF/DAOPHOT was performed.
For deriving PSFs, we selected unsaturated bright stars where the
highest pixel count was below the nonlinear sensitivity
regime (20,000 ADU) that were not close to the edge of
the frame and do not have any nearby stars with magnitude differences of
more than 5 mag.
The PSF photometry was performed using the ALLSTAR
routine with two iterations, once using the original images and a second
time using the images with sources from the first iteration subtracted.
We used PSF fit radii of 3.5, 3.75, and 3.75 pixels for the {\it J},
{\it H}, and {\it K} bands, which are the PSF FWHM values, and set the
inner radii and width of the sky annulus as four and three times as
large as the PSF fit radii, respectively.  Persson 9166 (GSPC P330-E;
$J=11.772$, $H=11.455$, and $K=11.419$ mag), which is an MKO standard
\citep{Leggett2006}, was used for the photometric calibration.  The
limiting magnitudes (10$\sigma$) based on the pixel-to-pixel noise of
long-exposure images for the S127 frame are $J=22.0$, $H=21.2$, and
$K_S=21.3$ mag, while those for the sky frame are $J=22.2$, $H=21.2$,
and $K_S= 21.0$ mag (see Table~\ref{tab:limit}).  The comparable
limiting magnitudes for the sky frame to those for the S127 frame
despite the shorter exposure times are probably due to the nebulosity of
S127.

Bright stars with magnitudes of $J \lesssim 17$, $H \lesssim 15.5$, and
$K_S \lesssim 15$ mag are saturated in long-exposure images.  For the
photometry of such bright stars, the short-exposure images are used.
The photometry was performed with the same procedure as for
long-exposure images but using a PSF fit radii of 7 pixels, which is the
PSF FWHM value, and setting the inner radii and width of the sky annulus
to four and three times as large as the PSF fit radii, respectively.
For the photometric calibration, stars whose magnitudes can be estimated
in both short- and long-exposure images with small uncertainties
(magnitudes of $J < 18.5$, $H < 17$, and $K_S < 17$ mag, and magnitude
uncertainties of $<$0.05 mag) were used.  For our analysis, we
preferentially adopt infrared luminosities from photometry with
long-exposure images because they have higher sensitivities and angular
resolution.  Very bright stars with magnitudes of $J \lesssim 13$, $H
\lesssim 14$, and $K_S \lesssim 12.5$ mag are saturated even in
short-exposure images in both the S127 and sky frames.

\section{Two young embedded clusters in S127} \label{sec:results} 

\subsection{Identification of Young Clusters in S127}
\label{sec:identify_cl}

Using the pseudocolor image (Figure~\ref{fig:3col_S127}), enhancements
of the stellar density compared to the surrounding area were identified
in the center of the field observed with MOIRCS.
The enhancements are located near the regions where the emission of {\it
WISE} band 3 (12 $\mu$m) is very strong
(Figure~\ref{fig:3col_2MASS_WISE}), which is often the case for young
clusters (see \citealt{Koenig2012}).

\begin{figure}[h]
\begin{center}
  \vspace{20em}
  \hspace{-10em}
  \includegraphics[width=19.5cm]{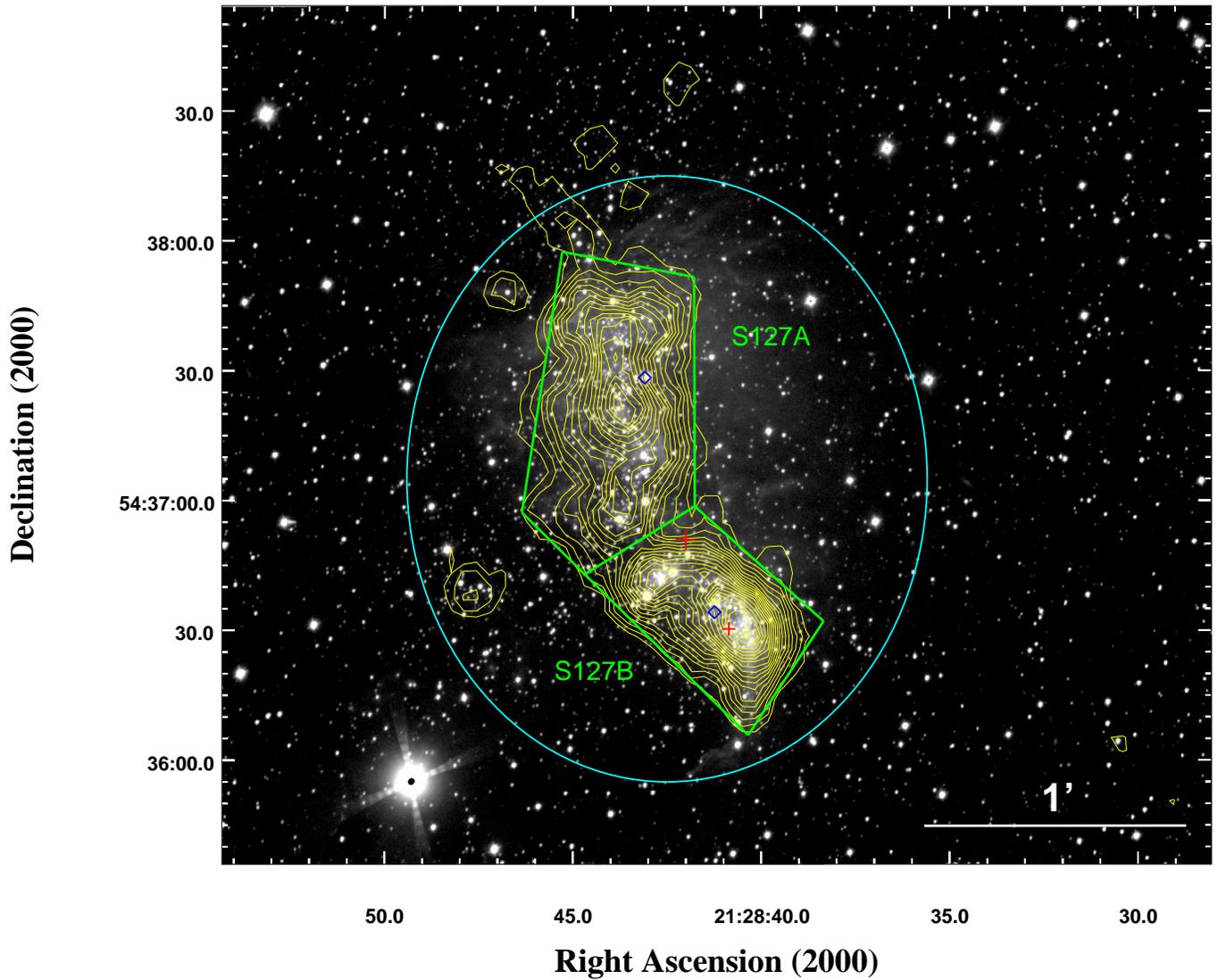}
  \vspace{-20em}
\caption{Stellar density of detected sources in the
MOIRCS $K_S$-band image is shown with yellow contours, 
superposed on the MOIRCS $K_S$-band image, whose field of view is the
same as Figure~\ref{fig:3col_S127}.  The contour levels represent
stellar densities of 3$\sigma$, 4$\sigma$, 5$\sigma$,
..., and 20$\sigma$ higher than the average stellar density in the field
outside of the cyan ellipse.
The green polygons show two identified star-forming clusters, S127A and
S127B clusters.}
\label{fig:SD}
\end{center}
\end{figure}

We determined the contour map of stellar density in the frame by
counting the number of stellar sources detected in the $K_S$ band that
are included within circles within a 50 pixel ($\sim$6$''$) radius. The
circles are set all over the frame with 25 pixel steps.
Because the stellar density is very high in
the center of the frame, we derived the background level from outside of
the cluster area, shown with a cyan ellipse in 
Figure~\ref{fig:SD}. 
The average number of stars in each of the circles is $13.8 \pm 6.1$.
Figure~\ref{fig:SD} shows the distribution of detected sources in the
MOIRCS $K_S$-band image with contour levels of 3$\sigma$, 4$\sigma$,
..., 20$\sigma$.  From the map, there are two large components that are
located in the north and south of the IRAS source.  The peak coordinates
are $\alpha_{\rm 2000} = 21^{\rm h} 28^{\rm m} 43.9^{\rm s}$,
$\delta_{\rm 2000} = +54^\circ 37' 21''$ and $\alpha_{\rm 2000} =
21^{\rm h} 28^{\rm m} 40.2^{\rm s}$, $\delta_{\rm 2000} = +54^\circ 36'
29''$ with an accuracy of $\sim$10$''$.  Both stellar enhancement peaks
are located very close to the peaks of the NVSS radio sources, shown
with blue diamonds, and the distributions of the clusters are consistent
with the distribution of \ion{H}{2} regions observed by
\citet{Rudolph1996} for both S127A and S127B
(Section~\ref{sec:SFinS127}).  The obscurations by molecular complex,
pointed out by \citet{Brand2001}, were seen in the northern half of the
S127A cluster and the west of the S127B cluster.
We defined the cluster regions, enclosed with green
polygons in the figure, as regions with stellar densities 3$\sigma$
larger compared to that of the entire frame.
The sizes of the two clusters are $\sim$$1\farcm2 \times 0\farcm8$ and
$\sim$$0\farcm8 \times 0\farcm6$ for S127A and S128B, respectively.
The size of the region including both clusters is $\sim$$2' \times 1'$,
which is consistent with the cluster size of angular dimensions of
$1\farcm9 \times 1\farcm1$ estimated in \citet{Bica2003}
with 2MASS data.
The size corresponds to $\sim$$6 \times 3$ pc with a distance of $D=10$
kpc.
We used the full sky frame as a control field.
The control field is used for subtracting the contamination of field
objects in the following discussion.

\subsection{Color--Magnitude Diagram} \label{sec:CM} 

\begin{figure*}
\begin{center}
  \vspace{10em}
  \hspace{-10em}
 \includegraphics[width=7cm]{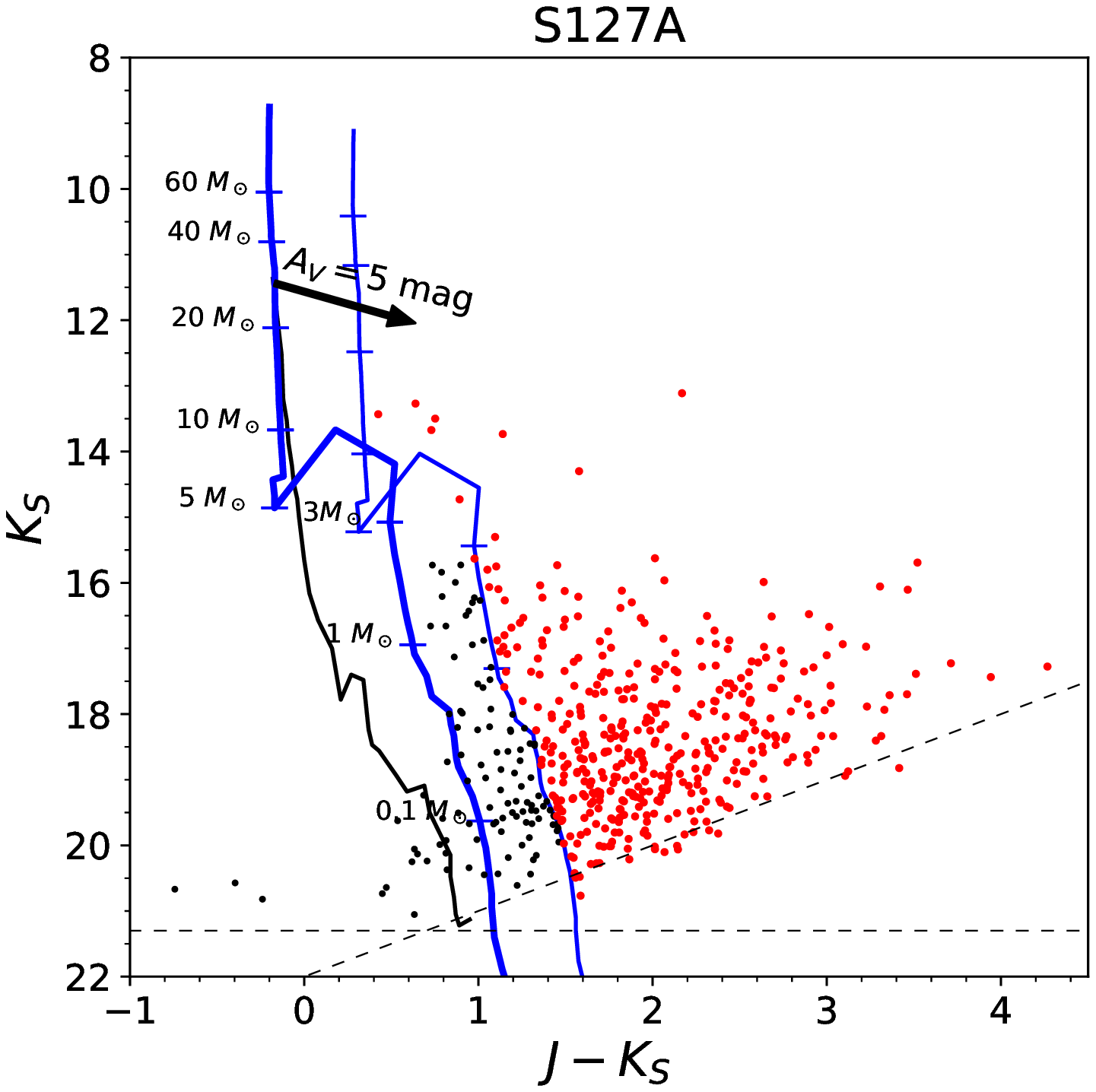}
\hspace{1em}
 \includegraphics[width=7cm]{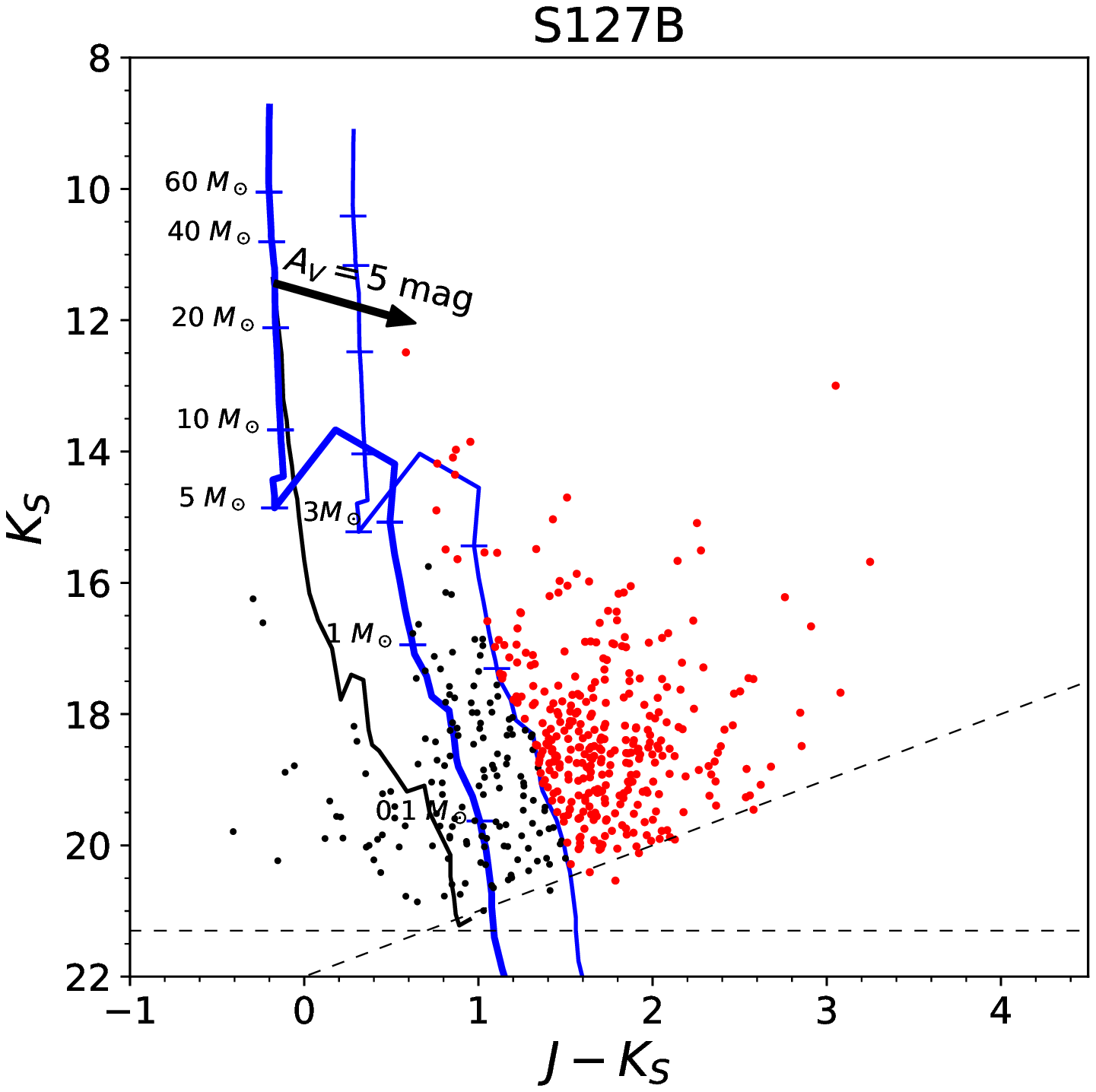}
  \vspace{-10em}
\caption{$(J - K_S)$ vs. {\it K}$_S$ color--magnitude diagram of the
 S127 clusters, the S127A (left) and S127B (right) cluster.  Identified
 cluster members in the cluster region ($A_V \ge 3$ mag) are shown with
 red dots, while other sources are shown with black dots.  The arrow
 shows the reddening vector of $A_V = 5$ mag.  The dashed lines mark the
 limiting magnitudes (10$\sigma$).  The black lines show the dwarf
 tracks by \citet{Bessell1988} in spectral types O9--M6 (corresponding
 mass of $\sim$0.1--20 $M_\odot$).  The blue lines denote the isochrone
 models for the age of 1 Myr by \citeauthor{D'Antona1997}
 (\citeyear{D'Antona1997}, \citeyear{D'Antona1998}; $0.017 \le M/M_\odot
 \le 3$), \citeauthor{Siess2000} (\citeyear{Siess2000}; $3 < M/M_\odot
 \le 7$), \citeauthor{Lejeune2001} (\citeyear{Lejeune2001}; $M/M_\odot
 \ge 7$).
The thick and thin lines show the isochrone models assuming $A_V=0$ and
3 mag, respectively.  A distance of 10 kpc is assumed.  The short
horizontal lines are placed on the isochrone models and shown with the
same colors as the isochrone tracks, which show the positions of 0.1, 1,
3, 5, 10, 20, 40, and 60 $M_\odot$.}  \label{fig:CM_JKK}
\end{center}
\end{figure*}

\begin{figure*}[!h]
\begin{center}
  \vspace{10em}
  \hspace{-10em}
 \includegraphics[width=7cm]{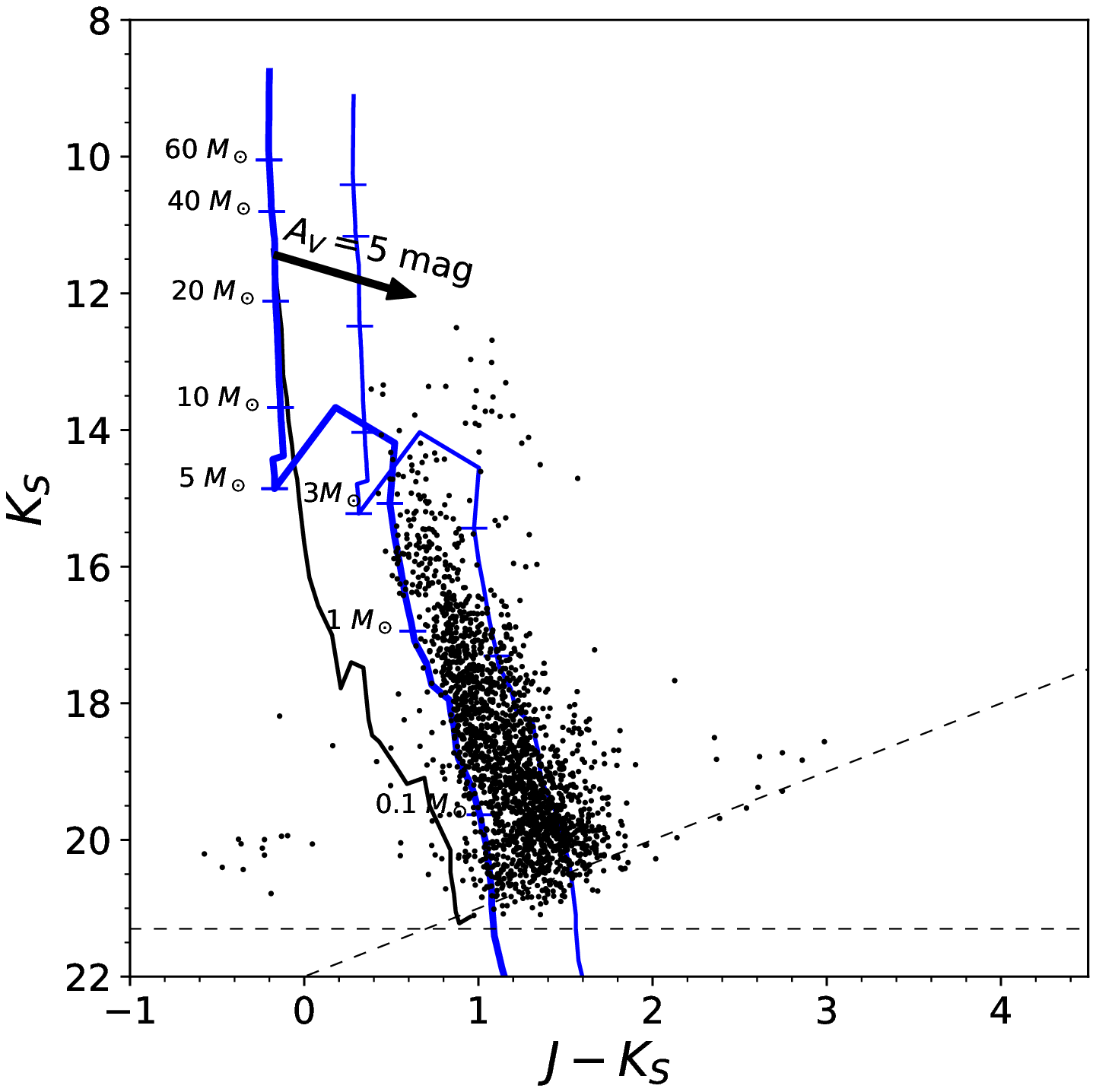}
  \vspace{-10em}
\caption{Same as Figure~\ref{fig:CM_JKK} but for the control field.}
\label{fig:CM_JKK_con}
\end{center}
\end{figure*}

\begin{figure*}[!h]
  \vspace{10em}
  \hspace{-10em}
\begin{center}
 \includegraphics[width=7cm]{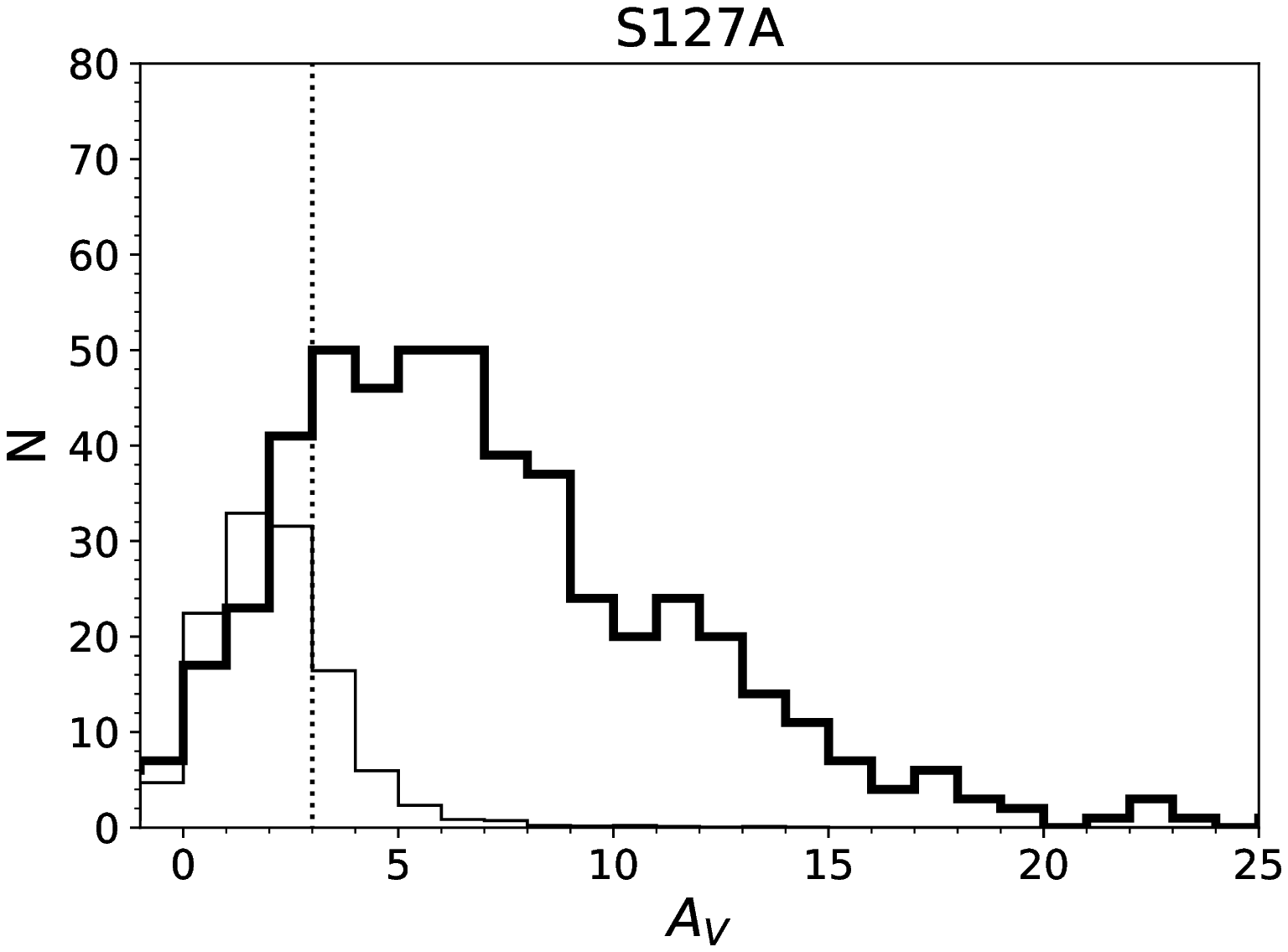}
\hspace{1em}
 \includegraphics[width=7cm]{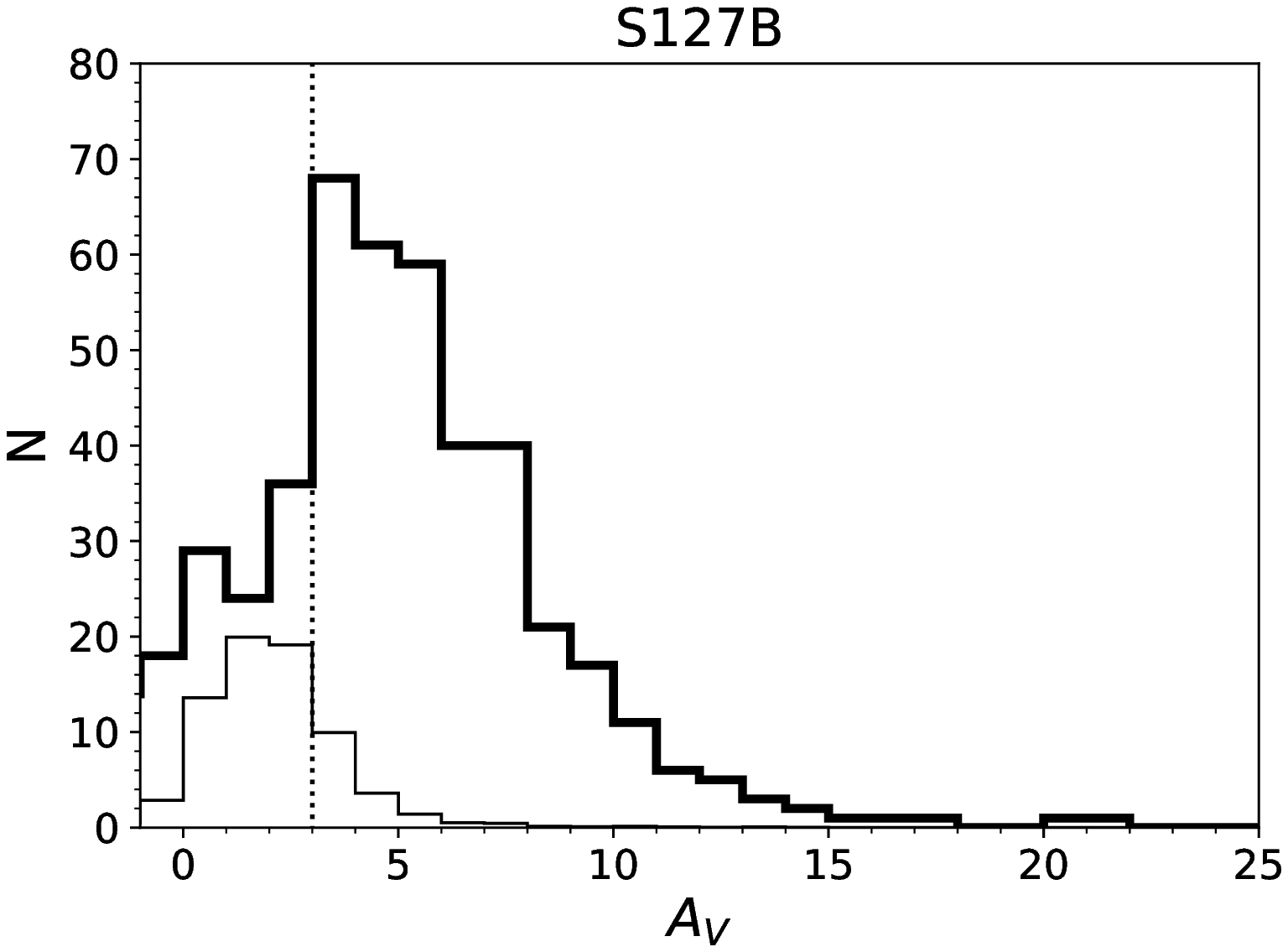}
  \vspace{-10em}
\caption{$A_V$ distributions for the sources in the S127 clusters (thick
 lines) and those in the control field (thin lines). Left and right
 panels show the S127A and S127B clusters, respectively.  The
 distributions for the control field are normalized to match the area of
 each cluster region.}  \label{fig:Av_CM}
\end{center}
\end{figure*}

For our S127 measurements, we followed the same investigation process as
described in Paper II for S208. Figure~\ref{fig:CM_JKK} shows the $J -
K_S$ versus $K_S$ color--magnitude diagrams for all detected point
sources in the S127A (left panel) and S127B (right panel) cluster
regions.  We also plotted detected point sources in the control field in
Figure~\ref{fig:CM_JKK_con}.  The isochrone models for the ages of 1 Myr
are shown as blue lines.  The models are by \citet{Lejeune2001} for the
mass of $ M/M_\odot \ge 7$, \citet{Siess2000} for the mass of $3 <
M/M_\odot \le 7$, and \citet{{D'Antona1997},{D'Antona1998}} for the mass
of $0.017 \le M/M_\odot \le 3$, and a distance of 10 kpc is assumed.
Arrows show the reddening vector of $A_V = 5$ mag.  In the
color--magnitude diagram, the extinction $A_V$ of each star was
estimated from the distance between its location and the $A_V=0$
isochrone models along the reddening vector.  Figure~\ref{fig:Av_CM}
shows the distributions of the extinction of stars in the cluster region
(thick lines) and control field (thin lines) for the S127A and S127B
clusters in the left and right panels, respectively.  The distribution
for the control field is normalized to match with the total area of the
cluster regions.  The resultant distribution for the control field shows
a peak of $A_V = 1$--3 mag, and then the number count decreases with
larger $A_V$, whereas that for the cluster region shows a peak at the
much larger extinction of $A_V \sim 3$--7 and 3--4 mag for the S127A and
S127B clusters, respectively, and they continue up to $A_V \sim 20$ mag.
This suggests that stars with $A_V \geq 3.0$ mag are concentrated in the
cluster regions, while stars with $A_V < 3.0$ mag are widely distributed
over the observed field.  Therefore, based on the values of $A_V$,
cluster members can be distinguished from contaminating noncluster stars
that appear in the cluster region, as is the case with the S208 clusters
(Paper II).  The following criteria are applied to identify members of
the S127 clusters: the stars (1) are distributed in the cluster regions
and (2) have large $A_V$ excess compared with normal field stars
(extinction of $A_V \ge 3$ mag).
In Figure~\ref{fig:CM_JKK}, the identified cluster members are shown by
red dots, while all other sources in the cluster regions are shown by
black dots.  As a result, 413 and 338 sources ($N_{\rm cl}$) are
identified as S127A and S127B cluster members, respectively.  The
average $A_V$ value of the cluster members is estimated at $A_V = 8.3
\pm 4.1$ and $A_V = 6.3 \pm 2.8$ mag for the S127A and S127B clusters,
respectively.

Considering the relatively large $R_g$ of the S127 clusters ($R_g =
13.5$ kpc), we can assume that the contamination of background stars is
negligible and most of the field objects are foreground stars
(the spatial density of background
galaxies is negligible at our detection limits, $K=21.3$ mag). 
To quantify the contamination, we compared the $A_V$ distributions of
all the sources in the cluster regions and the field objects in the
control field (Figure~\ref{fig:Av_CM}).
Because the number of field objects in the control field decreases
significantly at $A_V \ge 3$ mag, most cluster members can be
distinguished from the field objects as red sources with $A_V \ge 3$
mag. The contamination by the foreground stars is estimated at 7\% and
5\% for the S127A and S127B cluster, respectively, by counting the
normalized number of field objects in the tail of the distribution at
$A_V \ge 3$ mag and dividing it with the total number of sources in the
cluster regions.  In contrast, there must be some cluster members at
$A_V < 3$ mag that missed our identification.
The fraction of cluster members missed is estimated by $(N'_{\rm cl} -
N'_{\rm fi}) / (N_{\rm cl} + (N'_{\rm cl} - N'_{\rm fi}))$,  
where $N'_{\rm fi}$ is the normalized number of field objects with $A_V < 3$
mag, $N'_{\rm cl}$ is the number of stars in the cluster region with
$A_V < 3$ mag that do not meet the classification threshold, 
and $N_{\rm cl}$ is the number of identified cluster members.
The fractions of cluster members missed for the S127A and S127B clusters
are estimated at 3 \% and 11 \%, respectively.

On the isochrone models in Figure~\ref{fig:CM_JKK}, the positions of
0.1, 1, 3, 5, 10, 20, 40, and 60 $M_\odot$ are shown with horizontal
lines.  With the average $A_V$ for all S127 cluster members of $\sim$7
mag, the {\it K}-band limiting magnitude of 21.3 mag (10$\sigma$) for an
age of 1 Myr corresponds to a mass of 0.2 $M_\odot$ assuming a distance
of $D=10$ kpc.
The mass detection limit is sufficiently low, close to the substellar
mass limit, which enables by KLF fitting to derive parameters describing
the IMF down to around the IMF peak (Section~\ref{sec:IMF_age}) and to
derive the disk fraction with the same criteria as in the solar
neighborhood (Section~\ref{sec:DF}).
Because the most likely age of the S127 cluster is estimated at
$\sim$0.5 Myr in Section~\ref{sec:IMF_age}, the mass detection limit is
actually $\sim$0.2 $M_\odot$ with the average $A_V$ 
of $\sim$ 7 mag for all S127 cluster members.

\subsection{Color--Color diagram}\label{sec:CC} 

We constructed $J-H$ versus $H-K_S$ color--color diagrams for stars in
the S127 cluster regions (Figure~\ref{fig:CC_cl}).  Cluster members
identified in Section~\ref{sec:CM} are shown in red, while sources in
the cluster regions but not identified as cluster members are shown in
black.  We also constructed the color--color diagram for stars in the
control field (Figure~\ref{fig:CC_con}).  All sources that are detected
at more than 10$\sigma$ in all {\it JHK} bands are
plotted.  The dwarf star track for spectral types from late B to M6 in
the MKO system by \citet{Yasui2008} is shown with the blue curve.
The classical T Tauri star (CTTS) locus, originally derived by
\citet{Meyer1997} in the CIT system, is shown as a cyan line in the MKO
system \citep{Yasui2008}.
The arrow shows the reddening vector of $A_V = 5$ mag.

Stars in star-forming regions sometimes show large $H-K$ color excesses
due to their circumstellar dust disks, in addition to large extinctions
(e.g., \citealt{Lada1992}).
We estimated the color excesses for each source using the procedure 
described in Papers I and II. 
First, the intrinsic $(H-K)$ colors ($(H-K)_0$) were estimated by
dereddening along the reddening vector to the young star locus in the
color--color diagram (see Figure~\ref{fig:CC_cl}), which was
approximated by the extension of the CTTS locus.  Only stars that are
above the CTTS locus were used.
The obtained $(H-K)_0$ distributions for the S127 cluster members and
those in the control field are shown in Figure~\ref{fig:HK0_S127}.
After normalizing the distribution for the control field to the total
area of the cluster regions, there is a larger number of red stars with
$(H-K)_0 > 0.2$ mag in the distribution of the cluster members compared
to those in the control field.
The average $(H-K)_0$ values for cluster members are estimated at 0.39
mag for both the S127A and S127B clusters, whereas that in the control
field is estimated at 0.14 mag.
The difference in the average $(H-K)_0$ between the stars in the cluster
region and the field stars ($\simeq$0.25 mag) can be attributed to
thermal emissions from the circumstellar disks of the cluster members.
Therefore, we estimated the disk color excess of the S127 cluster
members in the {\it K} band, $\Delta K_{\rm disk}$, as 0.25 mag,
assuming that disk emissions appear in the $K$ but not in the {\it H}
band.

\subsection{KLF of the S127 clusters}
\label{sec:KLF} 

The KLF for the S127 cluster members is shown in Figure~\ref{fig:KLFobs}
as a black line up to the $K = 20.5$ mag bin.  The number counts of the
KLFs are relatively constant from $K=13.5$ and 12.5 mag for the S127A
and S127B clusters, respectively, and then increase toward the fainter
magnitude bins, with peaks of $K = 18.5$ mag for both clusters, and then
they decrease.
Although the general characteristics are the same for both clusters, the
slope in the brighter magnitudes for S127A is steeper than that for the
S127B cluster.  Because the 10$\sigma$ detection magnitude for the S127
frame is $K = 21.3$ mag, detection completeness should be $\sim$1 in all
magnitude bins (Section~\ref{sec:Data}; see also \citealt{Yasui2008} and
\citealt{Minowa2005}); therefore, the peak of the S127 KLFs at $K =
18.5$ mag would not change.

Because of the large $A_V$ dispersion of the S127 clusters ($A_V \sim
3$--20 mag), detection of faint cluster members may be difficult, which
could be the cause for the decrease of the KLF in fainter magnitude
bins.  For comparison, we also constructed KLFs for stars with limited
$A_V$ values in Figure~\ref{fig:KLFobs}, $A_V = 4.2$--12.4 and 3.5--9.1
mag (from the $A_V$ distribution of the cluster members, $8.3 \pm 4.1$
and $6.3 \pm 2.8$ mag for the S127A and S127B clusters, respectively;
Section~\ref{sec:CM}), which are shown with gray lines.
For clarity, the KLFs for limited $A_V$ samples are vertically shifted
by $+$0.1 mag for both clusters.  As a result, the discrepancy between
the KLFs for all cluster members and those for stars with limited $A_V$
values is found to be within the uncertainty range, suggesting that the
selection of stars with different limited $A_V$ values causes a
negligible influence on the obtained KLF.  Therefore, we used the
original KLF (the KLF from all S127 cluster members) in the following
discussion.

\section{Discussion} \label{sec:Discussion}

\subsection{Scale of the clusters} \label{sec:scale_cl}

\citet{Adams2006} found a clear correlation between cluster size and the
number of cluster members for young clusters in the solar neighborhood
from their embedded stage up to ages of $\sim$10 Myr.
Figure~\ref{fig:cl_scale} shows the number of stars in a cluster versus
cluster radius by open squares from the compilation of clusters in
\citet{LadaLada2003} and \citet{Carpenter2000}.
The figure shows that most clusters have $\sim$10--500 cluster members
($N_{\rm stars}$) and radii ($R$) of $\sim$0.2--2 pc.  \citet{Adams2006}
pointed out that the data can be fit by a relation of the form $R(N_{\rm
stars}) = R_{300} \sqrt{N_{\rm stars} /100}$ with $R_{300} = \sqrt{3}$
pc, shown in Figure~\ref{fig:cl_scale} with a solid line, and most data
points are scattered within a factor of $\sqrt{3}$ of $R_{300}$, shown
with dotted lines.

In Section~\ref{sec:results}, we identified two clusters in S127, the
S127A and S127B clusters.
The S127A cluster has 413 cluster members in a region of $\sim$$1\farcm2
\times 0\farcm8$, corresponding to a cluster radius of $\sim$$0\farcm6$,
while the S127B cluster has 338 cluster members in $\sim$$0\farcm8
\times 0\farcm6$, corresponding to a cluster radius of $\sim$$0\farcm4$.
Because 1$\arcmin$ corresponds to 3 pc at the 10 kpc distance of S127,
the cluster radii of the S127A and S127B clusters correspond to 1.8 and
1.2 pc, respectively.
We plot the values in Figure~\ref{fig:cl_scale} with red filled circles.
The plots show that the density of the S127B cluster is a little higher
than that of the S127A cluster.
Both clusters have both $R$ and $N_{\rm stars}$ that are within the
range found for clusters in the solar neighborhood but at the upper end
of that range.

We also plotted properties for other young low-metallicity clusters
(Section~\ref{sec:intro}) with red open circles: S207, S208, and the
Cloud 2-N and -S clusters (Papers I, II; \citealt{Yasui2008}).
They are relatively small clusters with $N_{\rm stars}$ of less than
100, even though the mass detection limits are similar to the limit for
S127 of $\lesssim$0.2$M_\odot$.
The $N_{\rm stars}$ is 73, 89, 52, and 59 for S207, S208, Cloud 2-N, and
Cloud 2-S, respectively, while the detection limits are
$\lesssim$0.2$M_\odot$ for S207, $\sim$0.2$M_\odot$ for S208, and
$\sim$0.1$M_\odot$ for Cloud 2-N and -S.
The cluster radii for all clusters are $\sim$1 pc: 1.3, 0.6, 1.4, and
0.6 pc for the S207, S208, Cloud 2-N, and Cloud 2-S clusters,
respectively.
The radii ($R$) of the S127 clusters are comparable to the radii of
those clusters.  However, the numbers of cluster members ($N_{\rm
stars}$) of the S127 clusters are larger by about an order of magnitude,
compared to other young low-metallicity clusters.

In the following sections, we estimate the underlying IMF and disk
fractions of the S127 clusters.
Because the S127 clusters are the first large-scale targets in
low-metallicity environments detected down to the substellar mass
regime, they are very appropriate targets to examine whether results
from previous young low-metallicity clusters hold even in such large
clusters.
If distinguished between metallicity dependence and dependence on
cluster scales (size and number of members), genuine metallicity
dependencies can be derived.
In fact, the possibility that the IMF and disk fraction depend on
cluster scales is suggested for clusters in the solar neighborhood,
i.e. a dependence on cluster mass and cluster density
 (see Sections~\ref{sec:IMF_age} and \ref{sec:DF}).
The S127 clusters are also useful for examining whether suggested
dependencies on cluster scales seen for clusters in the solar
neighborhood hold for clusters in low-metallicity environments.

\subsection{Implication for the IMF and age} \label{sec:IMF_age}

Although our final goal is to derive the IMF in low-metallicity
environments, information about the age of the cluster is necessary for
the derivation.
Here we estimate ages by assuming the canonical IMF observed in the
solar neighborhood as a first step, examine the adequacy of the
estimated age, and finally develop constraints on the underlying IMF.
We performed fitting of the KLF, which is known to strongly depend on
age and IMF, in the same way as Papers I and II. 
We note that we use observed KLFs in Section~\ref{sec:KLF} for the
fitting. 
Although it would be ideal if all of the detected sources were corrected
for extinctions derived in Section~\ref{sec:CM}, this cannot be possible
because not all stars are detected at more than one band, and at least
two bands of data are necessary for the derivation.
Moreover, especially for clusters with large extinctions, such as S208,
the longer NIR wavelength can detect lower mass stars due to the smaller
influence of extinctions, and that observed KLF is most appropriate for
deriving the IMF only from photometric data \citep{Muench2002}.
Therefore, we consider $A_V$ and $\Delta K_{\rm excess}$ values by
inputting them into model KLFs instead.  The method was originally
developed in \citet{Muench2002} and simplified in \citet{Yasui2006}.
We constructed model KLFs with ages from 0.1 to 3 Myr, which are shown
with colored lines in Figure~\ref{fig:KLFfit}.  A distance of 10 kpc is
assumed, and $A_V$ and $\Delta K_{\rm excess}$, which are estimated in
Sections~\ref{sec:CM} and \ref{sec:CC}, are applied ($A_V = 8.3$ and 6.3
mag for the S127A and S127B clusters, respectively, and $\Delta K_{\rm
excess} = 0.25$ mag).  The model KLF for 0.1 Myr has a shallower slope
of bright magnitudes before the peak than the KLF for 0.5 Myr.
The KLFs for 0.1--0.5 Myr have brighter peak magnitudes ($K=18.5$ mag)
than the KLFs for older ages ($K=19.5$ mag bin for 1 Myr and $K=20.5$
mag bin for 2--3 Myr).

The observed KLF for the S127A cluster is best fit with a model KLF for
an age of 0.5 Myr in terms of the slope for bright magnitudes and the
peak magnitude ($K=18.5$ mag bin), although the brightest observed
magnitude ($K=13.5$ mag bin) is a little off the model.
In contrast, for the S127B cluster, the slope of the observed KLF for
bright magnitudes is between that of the model KLFs for ages of 0.1 and
0.5 Myr, while the peak magnitude of the observed KLF matches those two
model KLFs.  After comparing with the KLF models, the ages of the S127A
and S127B clusters are estimated to be $\sim$0.5 and 0.1--0.5 Myr,
respectively.

The very young ages are consistent with the large $A_V$ values for S127A
and S127B cluster members estimated in Section~\ref{sec:CM}, up to
$\sim$20 mag with $A_V$ distributions of the cluster members of
$8.3 \pm 4.1$ mag for S127A and $6.3\pm 2.8$ mag for S127B.
\citet{Brand2001} reported that the column density of $^{12}$CO at the
peak position in S127B corresponds to $A_V \approx 8$ mag, which is
indeed consistent with our estimated values.
The distribution in the CO map by \citeauthor{Brand2001}
(\citeyear{Brand2001}; see their Figure~3(d)) shows that the majority of
the S127A cluster is embedded in molecular clouds.
Because the molecular gas disperses on a timescale of $\sim$3 Myr
\citep{Hartmann2001}, being fully embedded is consistent with the very
young age of the S127A cluster.
Although the S127B cluster is not fully embedded, it is located in a
compact \ion{H}{2} region (Section~\ref{sec:SFinS127}).
Compact \ion{H}{2} regions are thought to be a very early phase of
\ion{H}{2} regions, the next stage after the ultracompact \ion{H}{2}
phase \citep{Habing1979}.
Because the lifetime of ultracompact \ion{H}{2} regions is estimated to
be $\lesssim$0.1 Myr \citep{Hoare2007} and that of compact \ion{H}{2}
regions is estimated to be $\sim$0.3 Myr \citep{Mottram2011}, the
estimated age of the S127B cluster is consistent with these estimates.
The estimated age of the S127A cluster, which is located in a slightly
extended \ion{H}{2} region with a diameter of 1.7 pc, is older than the
estimated age of the S127B cluster located in a compact \ion{H}{2}
region with a diameter of 0.5 pc. This is consistent with older
\ion{H}{2} regions having larger radii \citep{Dyson1980}.
In addition, this is consistent with the results for S207 and S208 in
Papers I and II: the age of the S207 cluster located in an \ion{H}{2}
region with a diameter of 2.6 pc is estimated to be $\sim$2--3 Myr,
while that of the S208 cluster located in an \ion{H}{2} region with a
diameter of 1.4 pc is estimated to be $\sim$0.5 Myr.

From the KLF fitting using the typical IMF obtained in clusters with
solar metallicity, the S127 clusters are estimated to be very young
($\sim$0.5 Myr).  The estimated age is consistent with the independent
indications based on the very high H$_2$ column density and the size of
the \ion{H}{2} regions.  This suggests that the IMF of the S127
clusters, which are in a low-metallicity environment, is consistent with
the typical IMF in solar metallicity regions for masses $\ge$0.2
$M_\odot$.
Because the KLF slope for bright stellar magnitudes before the peak
strongly depends on the slope of the higher-mass region of the IMF
\citep{{Muench2000}, {Muench2002}}, the very good fit to the KLF peak
also suggests that the higher-mass IMF slope in the S127 clusters is
consistent with the typical IMF.
As for the KLF peaks, their magnitudes of $K=18.5$ mag for an age of 0.5
Myr correspond to a stellar mass of $\simeq$0.5 $M_\odot$. This is also
consistent with the canonical IMF within the margin of error, $\log M_c
/ M_\odot \sim 0.5 \pm 0.5$ \citep{Elmegreen2008}.
This is also the case for previous stuides of low-metallicity young
clusters, S207, S208, Cloud 2-N, and Cloud 2-S (Paper I, II;
\citealt{Yasui2008}; see also \citealt{Yasui2008ASPC},
\citealt{Yasui2017}).  Even with the similar mass detection limit for
S127, these other clusters are very small, with less than 100 cluster
members, while the S127 clusters have about an order of magnitude higher
numbers of cluster members, $\sim$300--400 (Section~\ref{sec:scale_cl}).

Note that in the case that detection is sufficiently deep, e.g., down to
the substellar mass regime ($\sim$0.1 $M_\odot$), the cluster mass is
roughly estimated on the very simple assumption that all stars have a
mass of 1 $M_\odot$ \citep[e.g.,][]{Yasui2008}.
With this assumption, the cluster masses of the S127 clusters are
estimated as $\sim$300--400 $M_\odot$, while those for the other young
low-metalicity clusters are $<$100 $M_\odot$.
Here we suggested that the IMFs for clusters with low-metallicity
environments are consistent with the typical IMFs observed in clusters
with solar metallicity regardless of cluster mass between $<$100 and
$\sim$400 $M_\odot$.
This is in the same manner as for clusters with solar metallicity
environments; there seem to be no clear indications that the IMF for
clusters with $\le$10$^3$ $M_\odot$ depends on their cluster scale,
while the possibility has been pointed out that the IMF for starburst
clusters, with $>$10$^4$ $M_\odot$, can change from the typical IMF
\citep{Bastian2010}.

\subsection{Disk Fraction} \label{sec:DF} 

The disk fraction is the fraction of cluster member stars with
protoplanetary disks out of all cluster members.  It is often used for
estimating disk lifetime, which is thought to be directly connected to
the duration of planet formation
\citep{{Haisch2001ApJL},{LadaLada2003}}.  Disk lifetime is estimated
using the age--disk fraction plot for various star-forming clusters
(Figure~\ref{fig:DF_age}; \citealt{Lada1999},
\citealt{Hillenbrand2005}).
In Figure~\ref{fig:DF_age}, we plot data for clusters in the solar
neighborhood as black squares \citep{{Yasui2009},{Yasui2010}}.  The fit
shown with a black curve is from \citet{Yasui2014}.
Disk fractions derived with NIR {\it JHK}-band observations show very
high values for very young clusters ($\sim$60 \%) that decrease with
increasing age.
The disk lifetime is often defined as the time when the disk fraction
reaches $\sim$5--10\%. 
Although NIR disk fractions are generally slightly lower than MIR disk
fractions \citep{Haisch2000} derived from ground based {\it L}-band
observations and space MIR observations, the characteristics are quite
similar (\citealt{Lada1999}, \citealt{Yasui2009}; see the red line in
the right panel of Figure~5 in \citealt{Yasui2014}).

\begin{figure}[!h]
\begin{center}
  \vspace{20em}
  \hspace{-10em}
 \includegraphics[width=8cm]{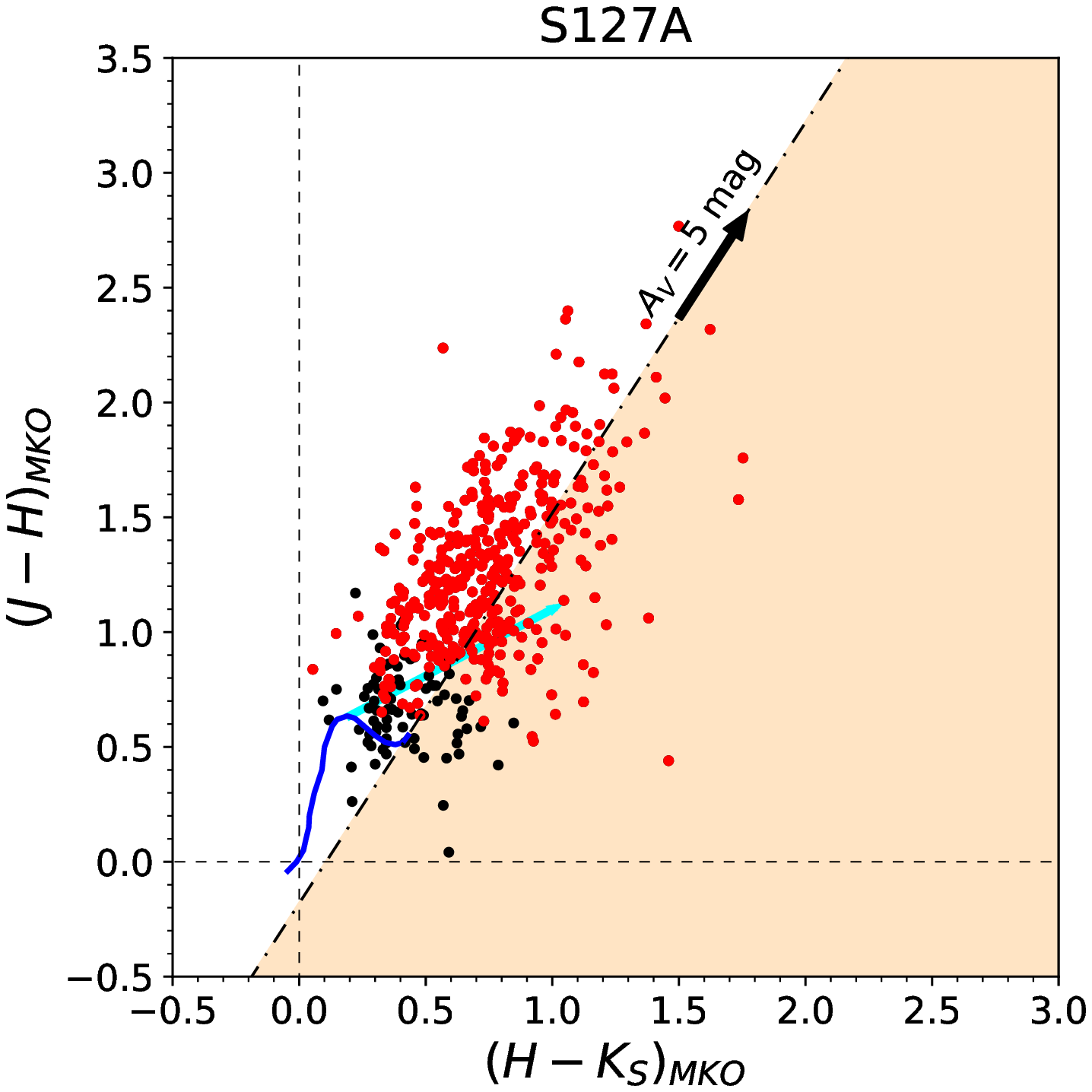}
\hspace{1em}
\includegraphics[width=8cm]{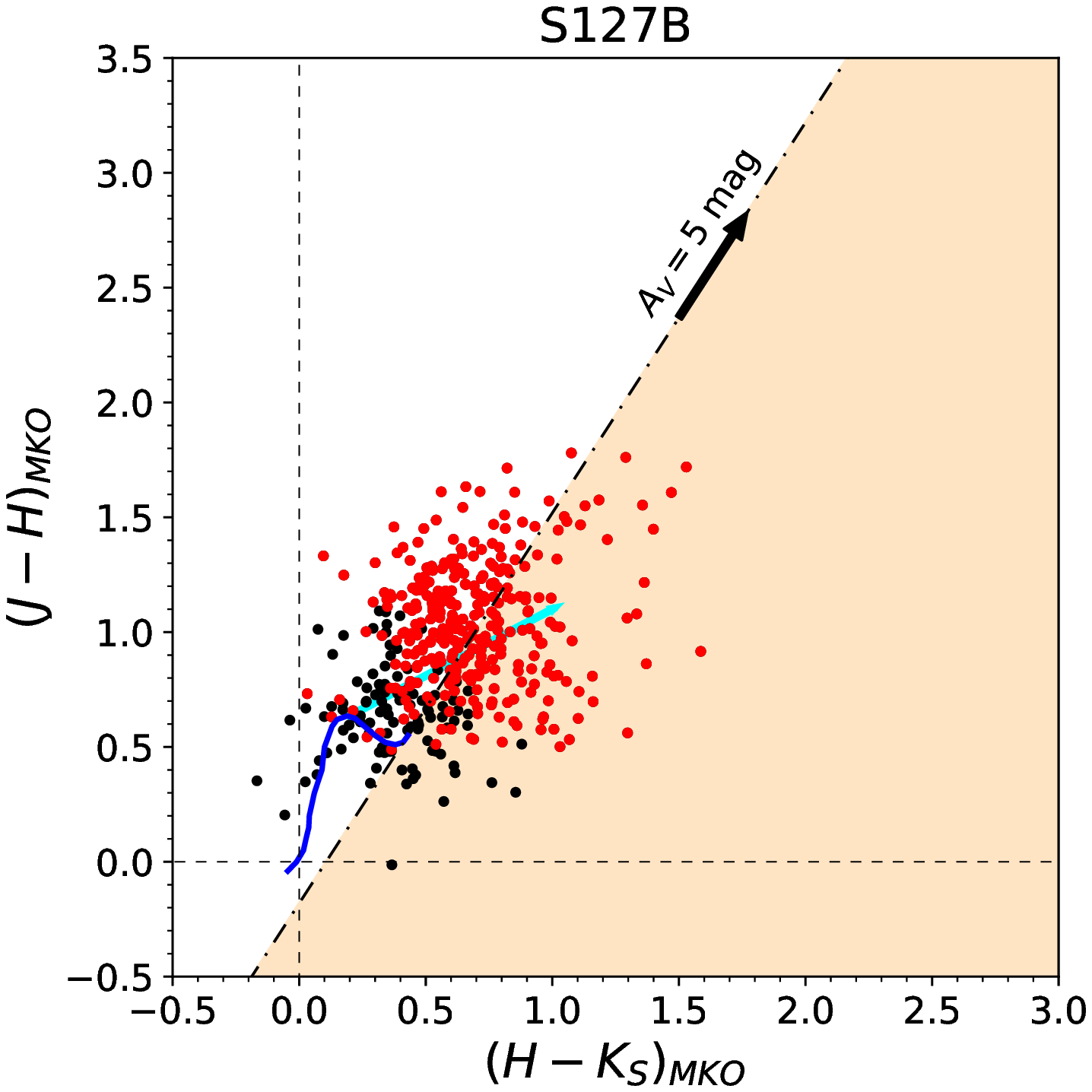}
\vspace{-10em}
\caption{$(H-K_S)$ vs. $(J - H)$ color--color diagram of S127.
Identified cluster members are shown in red, while sources in the
cluster region but not identified as cluster members are shown in black.
The blue curve in the lower left portion is the locus of points
corresponding to the unreddened main-sequence stars.  The dotted--dashed
line, which intersects the main-sequence curve at the maximum {\it
H}$-$$K_S$ values (M6 point on the curve) and is parallel to the
reddening vector, is the border between stars with and without
circumstellar disks.  The CTTS locus is shown with the
cyan line.}  \label{fig:CC_cl}
\end{center}
\end{figure}

\begin{figure}[!h]
\begin{center}
  \vspace{10em}
  \hspace{-10em}
 \includegraphics[width=8cm]{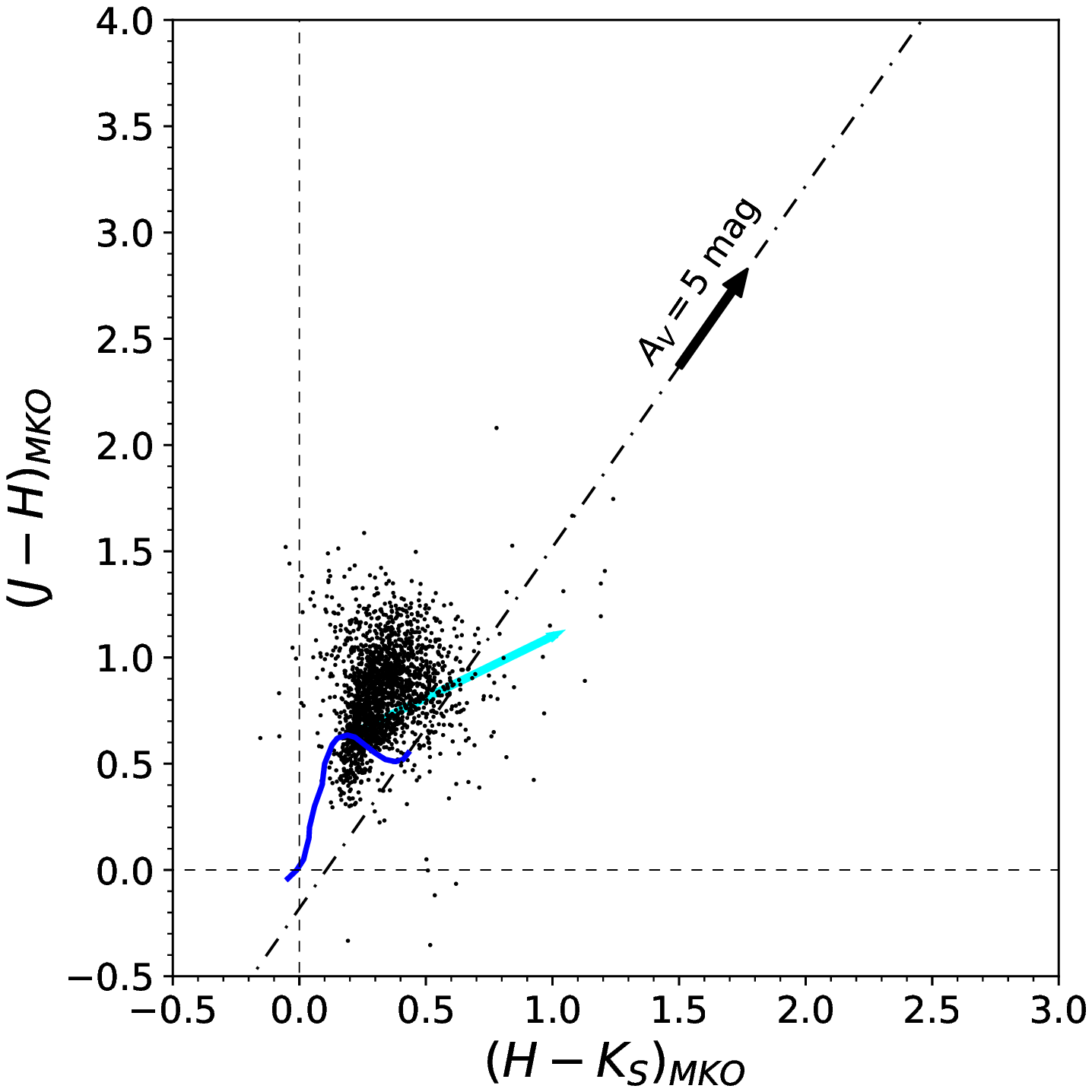}
  \vspace{-10em}
\caption{Same as Figure~\ref{fig:CC_cl} but for the control field.}
\label{fig:CC_con}
\end{center}
\end{figure}

We estimated the disk fraction for the S127 clusters using the NIR
color--color diagram (Figure~\ref{fig:CC_cl}) and the same method as
described in our previous papers.
In Figure~\ref{fig:CC_cl}, we used the dotted--dashed line parallel to
the reddening vector that passes through the point at the end of the
dwarf main-sequence star curve (blue line) at the point where the
$H-K_S$ value for the curve is maximum (the M6 point on the curve) as
the border between stars with and without circumstellar disks (see
details in \citealt{Yasui2009}).
Assuming that disk emission is only evident in the {\it K} band, we
classed stars on the lower right side of the borderline as disk excess
sources and calculated the ratio of the number of cluster members with
disk excesses to that of all cluster members.
As a result, disk fractions for the S127A and S127B clusters are
estimated to be $28\% \pm 3$\% $(108 / 391)$ and $40 \% \pm 4$\% $(128 /
318)$, respectively.
We plotted the disk fractions for the S127 clusters against their ages
estimated in Section~\ref{sec:IMF_age} in Figure~\ref{fig:DF_age} with
red filled circles with error bars.
Because the disk fractions are generally high in younger clusters, the
higher disk fraction for the S127B cluster than the S127A cluster is
consistent with the age estimates in
Section~\ref{sec:IMF_age}, where the S127B cluster is younger than the
S127A cluster.

\begin{figure}[!h]
\begin{center}
  \vspace{10em}
  \hspace{-10em}
 \includegraphics[width=8cm]{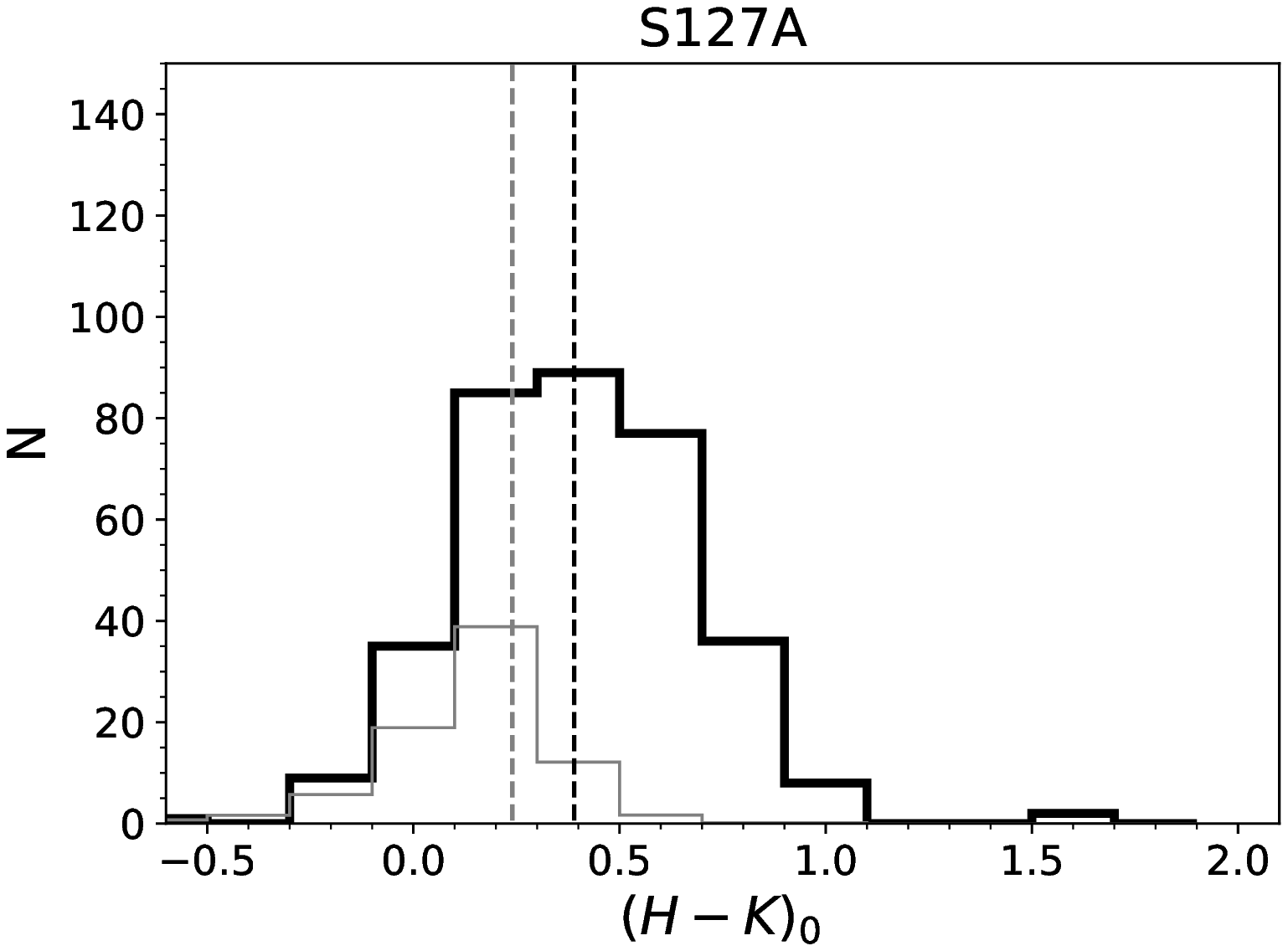} 
 \includegraphics[width=8cm]{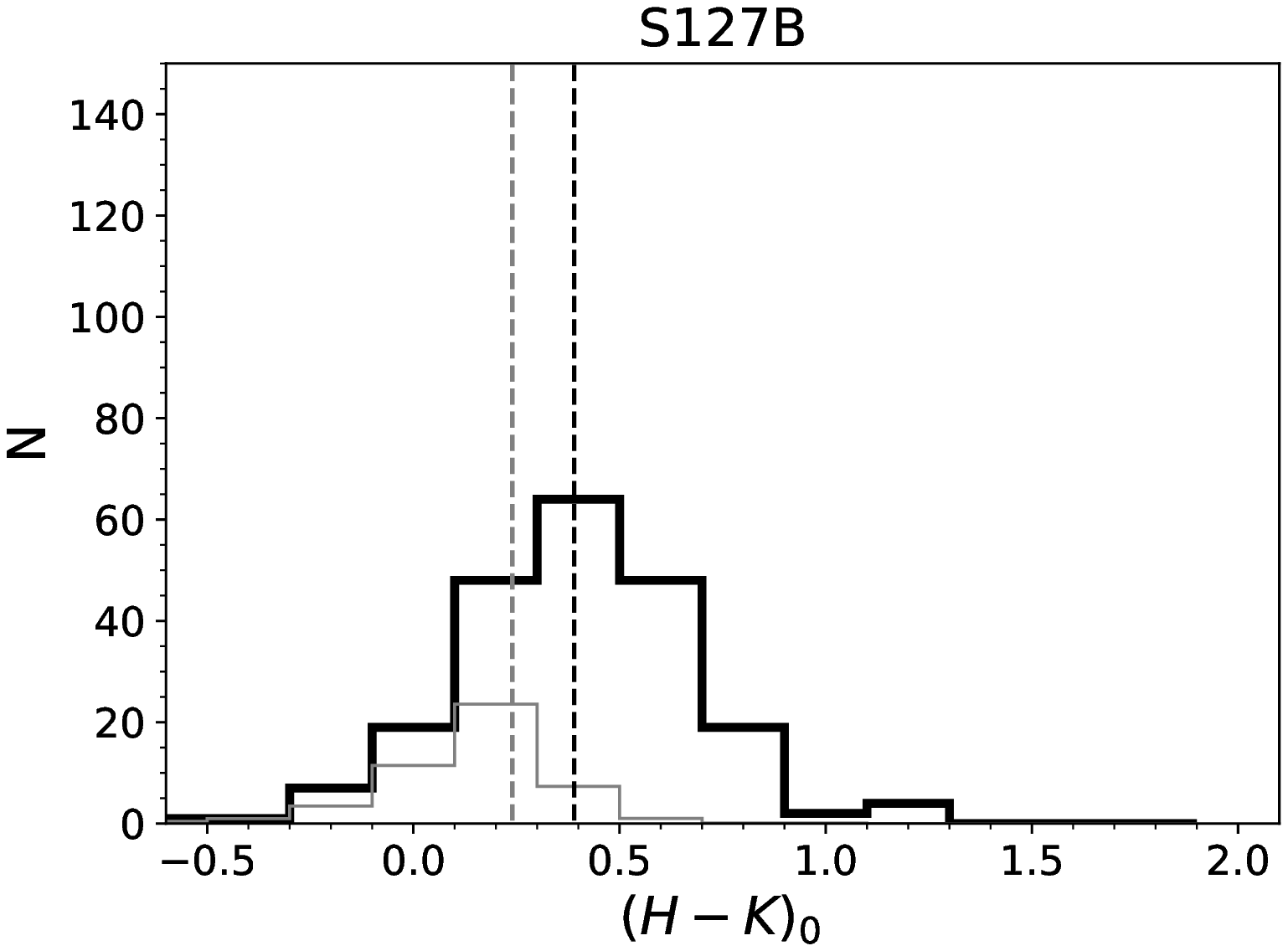} 
  \vspace{-10em}
\caption{$(H - K)_0$ distributions for the S127 cluster members (thick
line) and stars in the control field (thin line).  Left and right panels
show the S127A and S127B clusters, respectively.  The distribution for
the control field is normalized to match the total area of the cluster
region.
The vertical black and gray dashed lines show average $(H-K)_0$ values
for the S127 cluster members and stars in the control field, respectively. 
 }  \label{fig:HK0_S127}
\end{center}
\end{figure}

\begin{figure}[!h]
\begin{center}
  \vspace{10em}
  \hspace{-10em}
  \includegraphics[width=8.5cm]{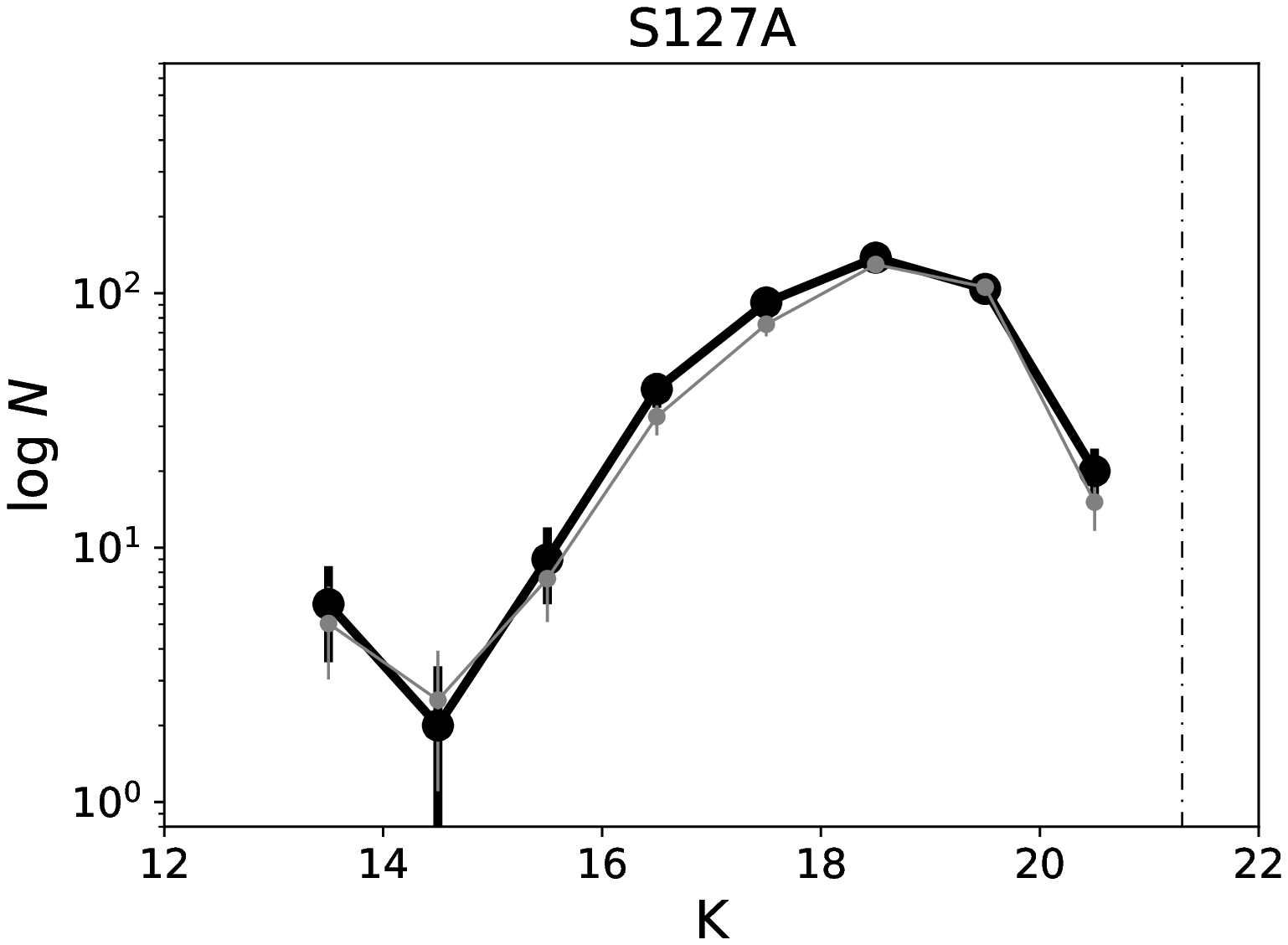}
\hspace{1em}
\includegraphics[width=8.5cm]{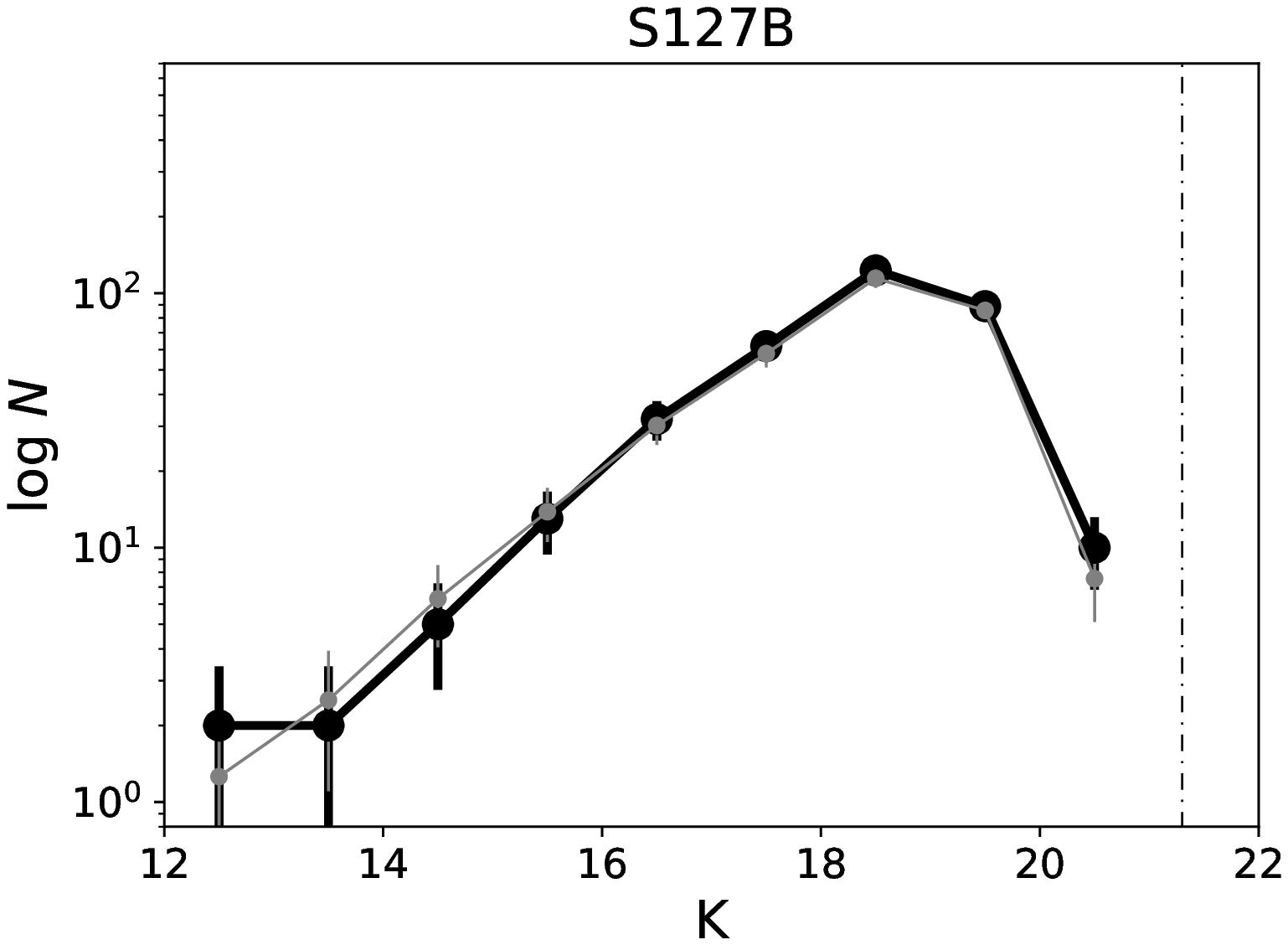}
\vspace{-10em}
\caption{Raw KLFs for the S127A cluster members (left) and S127B
(right).  The KLF for all cluster members with $A_V \ge 3$ mag is shown.
Error bars are the uncertainties from Poisson statistics.
The KLFs for limited $A_V$ samples are shown by gray lines for cluster
members with $A_V$ of 4.2--12.4 and $A_V$ of 3.5--9.1 mag for S127A and
S127B, respectively (see text for more detail).  For clarity, the KLFs
for limited $A_V$ samples are vertically shifted by +0.1 for both
clusters.  The vertical dotted--dashed line shows the limiting
magnitudes of the 10$\sigma$ detection (21.3 mag).}

\label{fig:KLFobs}
\end{center}
\end{figure}

\begin{figure}[!h]
\begin{center}
  \vspace{10em}
  \hspace{-10em}
  \includegraphics[width=8cm]{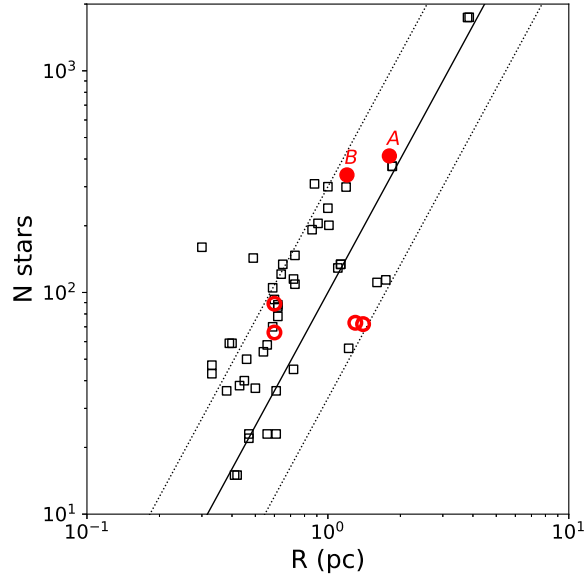}
  \vspace{-10em}
\caption{Correlation between the number of stars in a cluster ($N_{\rm
stars}$) and the radius of the cluster ($R$).
The red filled circles show S127 clusters,
 while red open circles show other young low-metallicity cluster
 samples: S207, S208, and Cloud 2-N and -S from other papers in our
 series.
The open squares show clusters in the solar neighborhood whose data are
from \citet{LadaLada2003} and \citet{Carpenter2000}.
The solid line shows a rough fit to the data for clusters in the solar
neighborhood; most points are scattered within a factor of $\sqrt{3}$ of
$R$, shown with dotted lines. The solid and dotted lines represent lines
of constant cluster density.}  \label{fig:cl_scale}
\end{center}
\end{figure}

\begin{figure}[!h]
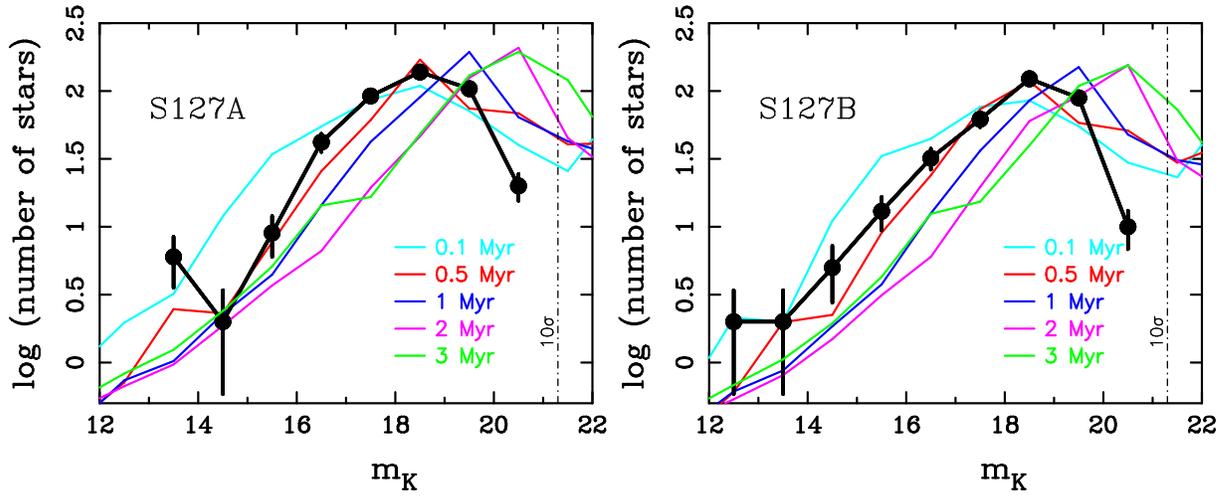

\begin{center}
 \includegraphics[width=8cm]{KLF_S127A_2020Aug.eps}
 \includegraphics[width=8cm]{KLF_S127B_2020Sep.eps}
\caption{Comparison of the S127 KLFs (black lines) with model KLFs of
various ages (colored lines).
Error bars are the uncertainties from Poisson statistics. 
The cyan, red, blue, magenta, and green lines represent model KLFs of
0.1, 0.5, 1, 2, and 3 Myr, respectively.
The vertical dotted--dashed lines show the limiting magnitudes of the
10$\sigma$ detection (18.5 mag).} \label{fig:KLFfit}
\end{center}
\end{figure}

\begin{figure}[!h]
\begin{center}
  \includegraphics[scale=0.5]{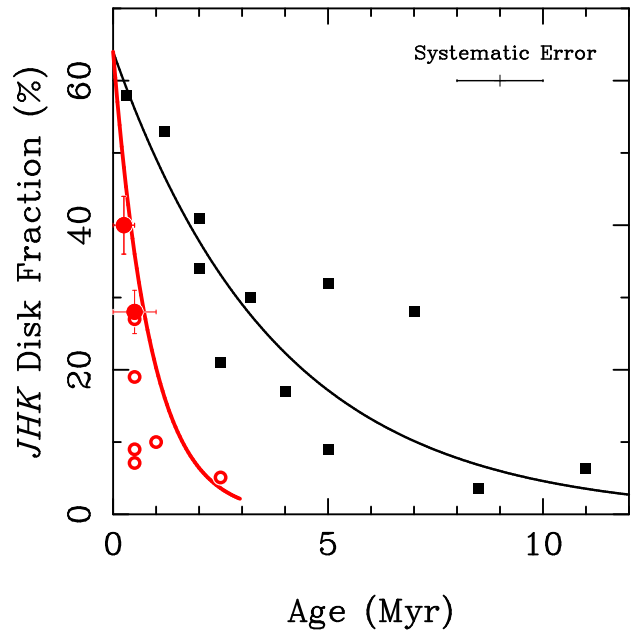}
\caption{Disk fraction as a function of cluster age.  {\it JHK} disk
fractions of the young clusters in low-metallicity environments are
shown with red circles.
The S127 clusters are shown with red filled circles, while other
clusters are shown with red open circles.  {\it JHK} disk fractions of
young clusters with solar metallicity are shown by black filled squares.
The black and red lines show the disk fraction evolution under solar
metallicity and in low-metallicity environments, respectively.}
\label{fig:DF_age}
\end{center}
\end{figure}

\begin{figure}[!h]
\begin{center}
  \includegraphics[scale=0.5]{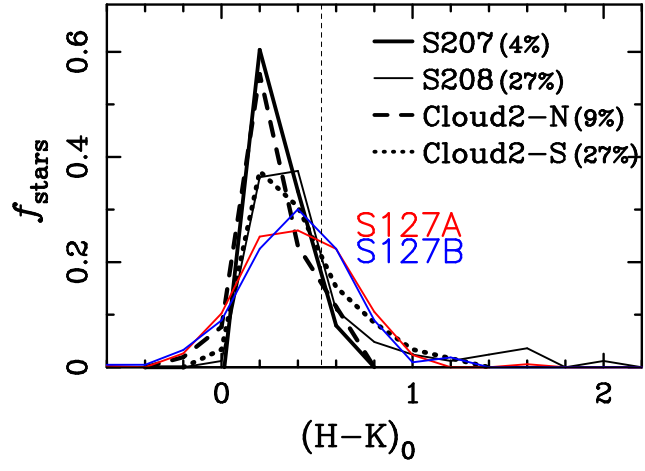}
\caption{Comparison of intrinsic $H-K$ color distributions. The
fractions of stars ($f_{\rm stars}$) per each intrinsic color bin
$(H-K)_0$ for clusters in low-metallicity environments, S208, S207,
Cloud 2-N, and Cloud 2-S, are plotted. The vertical dashed line shows the
borderline for estimating the disk fraction in the MKO system.}
\label{fig:HK0_disk}
\end{center}
\end{figure}

In Figure~\ref{fig:HK0_disk}, we show the fraction of stars ($f_{\rm
stars}$) in each intrinsic $(H-K)$ color bin $(H-K)_0$ for the S127A
cluster in red and the S127B cluster in blue (also shown as black lines
in Figure~\ref{fig:HK0_S127}), as well as those for other young clusters
in low-metallicity environments: S207 (thick solid line), S208 (thin
solid line), Cloud 2-N (dashed line), and Cloud 2-S (dotted line).
The vertical dashed line represents the borderline for estimating the
disk fraction in the MKO system, i.e. the dotted--dashed line in
Figure~\ref{fig:CC_cl}. 
The distribution becomes bluer and sharper with lower disk fractions for
nearby young clusters (see the bottom panel of Figure~7 in
\citealt{Yasui2009}), which is also the case for clusters in
low-metallicity environments \citep{Yasui2009}.  The peak $(H - K)_0$ of
the S127 cluster is relatively red, $(H - K)_0 \sim 0.4$ mag, and the
distribution is relatively broad, with a maximum $(H - K)_0$ of
$\sim$1.5 mag.
The distributions of the S127 clusters resemble those of the S208 and
Cloud 2-S clusters with disk fractions of $\sim$30\% rather than those
of the S207 and Cloud 2-N clusters with disk fractions of $<$10\%.
This suggests that the distributions of the S127 clusters are consistent
with the estimated disk fraction.

Considering the very young age of the S127 clusters, $\sim$0.5 Myr,
their disk fraction is significantly lower than that for clusters in the
solar neighborhood with the same age ($\sim$50--60 \%).  This is the
case for other young clusters in low-metallicity environments, the S207,
S208, and Cloud 2-N and -S clusters, which are shown with red open
circles in Figure~\ref{fig:DF_age}.  The data are from Paper I, Paper
II, and \citet{Yasui2009}, and we also plot data for the S209 clusters
in \citet{Yasui2010}.
We also show the fit for the clusters with a red curve, which is
obtained using the same procedure as for clusters in the solar
neighborhood \citep{Yasui2014}, assuming the same initial disk fraction
at $t=0$ as for clusters in the solar neighborhood of 64\%.
\citet{Yasui2010} pointed out the tendency for low-metallicity clusters
to have lower disk fractions than solar metallicity clusters for a given
age and suggested that the disk lifetime in low-metallicity environments
is quite short, as discussed in \citet{Yasui2009}.
However, note that the estimated disk fraction for the S127B cluster is
higher than those for other low-metallicity clusters.
It may be possible that the disk fraction for the S127B cluster is
elevated due to factors other than metallicity and young age (e.g.,
position in the Galaxy, cluster scale, etc).
However, the disk fraction for the S127A cluster is comparable to those
for clusters with the same age, S208 and Cloud 2-S, despite the fact
that S127A is in the same location in the Galaxy as S127B and both have
identical cluster scales.
Therefore, the high value of the disk fraction for the S127B cluster is
probably due to the very young age, between 0.1 and 0.5 Myr, which is
the youngest among the low-metallicity cluster sample.
The results that quite young clusters in 
low-metallicity environments have relatively high disk fractions, although
disk fractions of only $\lesssim$30 \% had been obtained for the
clusters previously observed, suggest that a large fraction of stars have
disks even in low-metallicity environments in the very early phases and that
they lose their disks on a very short timescale.

In the solar neighborhood, it is suggested that disk fraction, and thus
disk lifetime, depends on cluster scales, such as cluster mass and
stellar density \citep[e.g.,][]{{Fang2013}, {Stolte2010}}.
\citet{Fang2013} suggested that the disks in sparse stellar associations
are dissipated more slowly than those in denser (cluster) environments,
while \citet{Stolte2010} suggested that disk depletion is significantly
more rapid in compact starburst clusters than in moderate star-forming
environments.
In the previous studies of young clusters in low-metallicity
environments, observed clusters are relatively small, with numbers of
identified cluster members of less than 100, while the S127 clusters
have $\sim$300--400 cluster members, as discussed in
Section~\ref{sec:scale_cl}.
Although both cluster scales are common in the solar neighborhood, the
S127 clusters are the largest clusters in their class.  The result that
the disk fractions for the S127 clusters, with their larger number of
cluster members, are lower than those clusters in the solar neighborhood
with similar ages, which is also the case for smaller clusters, suggests
that a lower disk fraction for a given age is a characteristic of
clusters in low-metallicity environments.
However, because the number of low-metallicity clusters studied is small,
it is necessary to study more star-forming clusters in low-metallicity
environments and cover a range of cluster scales, e.g., cluster mass and age.

Finally, it should be noted that class I protostar candidates, which
were seen in the S208 clusters, are also seen in the S127 clusters: nine
sources in S127A and three sources in S127B.
The candidates are selected here using the same method as in Paper II,
i.e. sources having large $J-K$ colors of larger than 3 (equal to the
sum of $J-H$ (y-axis) and $H-K$ (x-axis) colors on the NIR color--color
diagram).
This suggests that young stellar objects in low-metallicity environments
are initially surrounded by thick circumstellar materials, as is the
case for the solar neighborhood, but they disperse very quickly, as also
discussed in Paper II.  This also supports that the very young age
estimated for S127 in Section~\ref{sec:IMF_age} (0.5 Myr) is reasonable,
considering the estimated age of S208 is also estimated as $\sim$0.5
Myr.


\acknowledgments

This work was supported by JSPS KAKENHI grant No.  26800094.
We thank the Subaru support staff, in particular MOIRCS support
astronomer Ichi Tanaka. We also thank Chihiro Tokoku for helpful
discussions on the observation.



\begin{table*}[h]
\caption{Properties of S127.}\label{tab:targets} 
\begin{center}
\begin{tabular}{llcccccccc}
\hline
\hline
Name & S127 \\
\hline
Galactic longitude (deg) & 96.287 (1) \\
Galactic latitude (deg) &  $+$2.594 (1) \\
R.A. (J2000.0) & 21 28 41.6   (1) \\
Decl. (J2000.0) & $+$54 37 00 (1) \\ 
Photometric heliocentric distance (kpc) 
 & 9.7 (2) \\
Kinematic heliocentric distance (kpc) &  9.97 (3)\\
Adopted distance (kpc) &  $\simeq$10 \\
Galactocentric distance$^{\rm a}$ (kpc) & $\simeq$13.5 \\
Oxygen abundance $12 + \log {\rm (O/H)}$ & 
$8.15^{+0.12}_{-0.17}$ (4, 5), $8.20^{+0.15}_{-0.23}$ (5, 6),
     $8.20^{+0.17}_{-0.29}$ (5, 7), 
     $7.68^{+0.17}_{-0.29}$ (5, 8) \\  
Metallicity [O/H] (dex)$^{\rm b}$ & $\simeq$$-$0.7 \\
Electron temperature (K) & 10500$\pm$820 (9), 11428$\pm$305 (10)\\
\hline
\end{tabular}
\end{center}
{{\small {\bf Notes.} References are shown in parentheses. \\
 $^{\rm a}$Assuming a solar Galactocentric distance
 of $R_\odot = 8.0$ kpc. \\
 $^{\rm b}$Assuming a solar abundance of $12+ \log {\rm
 (O/H)} = 8.73$ \citep{Asplund2009}. \\
{\small \bf References. }
(1) SIMBAD \citep{Wenger2000}, 
(2) \citet{Chini1984}, 
(3) \citet{Foster2015}, 
(4) \citet{Vilchez1996},
(5) \citet{Rudolph2006},
(6) \citet{Rudolph1997},
(7) \citet{Caplan2000}, 
(8) \citet{Peeters2002},
(9) \citet{Scaife2008}, 
(10) \citet{Balser2011}.}} 
\end{table*}

\begin{table*}[!h]
\caption{Summary of MOIRCS Observations.} \label{tab:LOG}
\begin{center}
\begin{tabular}{lcccccccc}
\hline
\hline
Modes & Date & Band & $t_{\rm total}$ & $t$ & Coadds & $N_{\rm total}$
 & Seeing & Sky Condition\\ 
(1) & (2) & (3) & (4) & (5) & (6) & (7) & (8) & (9)\\
\hline 
$J$-long & 2006 Sep 2 & $J$ & 1350 & 150 & 1 & 9 (4) & $0\farcs4$& P \\

$H$-long & 2006 Sep 2 & $H$  & 1080 & 20 & 6 & 9 (4) & $0\farcs4$ & P \\

$K_S$-long & 2006 Sep 2 & $K_S$ & 1080 & 30 & 4 & 9 (4) & $0\farcs4$ & P \\ 

$J$-short & 2007 Nov 22 & $J$  & 52  & 13 & 1 & 4 (3) & $0\farcs8$ &H \\
$H$-short & 2007 Nov 22 & $H$  & 52  & 13 & 1 & 4 (3) & $0\farcs8$ &H\\
$K_S$-short & 2007 Nov 22 & $K_S$  & 52 & 13 & 1 & 4 (3)  & $0\farcs8$ & H\\
\hline
\end{tabular}
\end{center}
{{\small {\bf Notes.} 
Column (4): total exposure time (s). 
Column (5): single-exposure time (s).
Column (6): number of coadds. 
Column (7): total number of frames. 
Column (9): P: photometric, and H: high humidity.  
The values for the sky frames are shown in parentheses.}}
\end{table*}

\begin{table*}[!h]
\caption{Limiting magnitudes (10$\sigma$) of long-exposure images for
MOIRCS observations.} \label{tab:limit}
\begin{center}
\begin{tabular}{lccc} 
\hline
\hline
Frame & $J$ Band & $H$ Band & $K_S$ Band \\
\hline
Cluster & 22.0 & 21.2 & 21.3 \\
Sky & 22.2 & 21.2 & 21.0 \\
\hline
\end{tabular}
\end{center}
\end{table*}

\end{document}